\documentclass[modern]{aastex62}

\usepackage{url}
\usepackage{amsmath}
\usepackage{mathtools}
\usepackage{amssymb}
\usepackage{natbib}
\usepackage{graphicx}
\usepackage{calc}
\usepackage{etoolbox}
\usepackage{xspace}
\usepackage[T1]{fontenc} % https://tex.stackexchange.com/a/166791
\usepackage{textcomp}
\usepackage{ifxetex}
\ifxetex
\usepackage{fontspec}
\defaultfontfeatures{Extension = .otf}
\fi
\usepackage{fontawesome}
\usepackage{listings}
\usepackage{nicefrac}
\usepackage[bb=boondox]{mathalfa}
\usepackage{booktabs}
\usepackage{longtable}

\renewcommand{\eqref}[1]{\ref{eq:#1}}

\definecolor{linkcolor}{rgb}{0.1216,0.4667,0.7059}
\newcommand{\codeicon}{{\color{linkcolor}\faFileCodeO}}
\newcommand{\prooficon}{{\color{linkcolor}\faPencilSquareO}}
\newcommand{\codelink}[1]{\href{https://github.com/fbartolic/volcano/blob/59146a0fa6a3a486049f29b01105abec15822e9d/paper/figures/#1.py}{\codeicon}\,\,}
\newcommand{\animlink}[1]{\href{https://github.com/fbartolic/volcano/blob/59146a0fa6a3a486049f29b01105abec15822e9d/paper/figures/#1.gif}{\animicon}\,\,}
\newcommand{\prooflink}[1]{\href{https://github.com/fbartolic/volcano/blob/59146a0fa6a3a486049f29b01105abec15822e9d/paper/proofs/#1.ipynb}{\raisebox{-0.1em}{\prooficon}}}
\newcommand{\cilink}[1]{\href{https://dev.azure.com/fbartolic/volcano/_build}{#1}}

\newtagform{eqtag}[]{(}{)}
\newcommand{\currentlabel}{None}

\usepackage[skins]{tcolorbox}
\newtcbox{\figtimebox}{enhanced,nobeforeafter,tcbox raise=-0.8mm,boxrule=0.6pt,
  top=0.5mm,bottom=0mm,right=0mm,left=6mm,arc=1pt,boxsep=2pt,
  before upper={\vphantom{dlg}},colframe=linkcolor,coltext=linkcolor,
  fontupper=\sffamily\bfseries\tiny,colback=white,overlay={\begin{tcbclipinterior}
  \fill[linkcolor] (frame.south west)
  rectangle node[text=white,font=\sffamily\bfseries\tiny,rotate=0]{CPU} 
  ([xshift=6mm]frame.north west);\end{tcbclipinterior}}}
\robustify{\figtimebox}
\newcommand{\figtime}[1]{\IfFileExists{figures/#1.py.time}%
{%
\cilink{\figtimebox{\input{figures/#1.py.time}\unskip s}}
}{}}

\newcommand{\oscaption}[2]{\caption{#2 \codelink{#1} \input{figures/#1.py.time}}}

\definecolor{codegreen}{rgb}{0,0.6,0}
\definecolor{codegray}{rgb}{0.5,0.5,0.5}
\definecolor{codepurple}{rgb}{0.58,0,0.82}
\definecolor{backcolour}{rgb}{0.95,0.95,0.95}
\lstdefinestyle{mystyle}{
    backgroundcolor=\color{backcolour},
    commentstyle=\color{codegreen},
    keywordstyle=\color{magenta},
    numberstyle=\tiny\color{codegray},
    stringstyle=\color{codepurple},
    basicstyle=\small\ttfamily,
    breakatwhitespace=false,
    breaklines=true,
    captionpos=b,
    keepspaces=true,
    numbers=left,
    numbersep=5pt,
    showspaces=false,
    showstringspaces=false,
    showtabs=false,
    tabsize=2,
    aboveskip=1em,
    belowskip=1em,
    keywords=[2]{map},
    keywordstyle=[2]{\color{black!80!black}},
    upquote=true
}
\renewcommand\quad{\hskip\fontdimen3\font}

\usepackage{stackengine,scalerel}
\def\lnlam{\ThisStyle{\ensurestackMath{\stackon[-2.4\LMpt]{%
  \SavedStyle\lambda}{\kern-.5pt\kern\LMpt\rule{1\LMex}{.25pt+.15\LMpt}}}}}

\begin{document}

% Title
\title{
    \vspace{-3em}
\textbf{Occultation mapping of Io's surface in the near-infrared \\ I: Inferring static maps} 
}
% Author list
\author[0000-0001-8630-9794]{Fran Bartoli\'c}
\email{fb90@st-andrews.ac.uk}
\affil{Centre~for~Exoplanet~Science, SUPA, School~of~Physics~and~Astronomy, University~of~St.~Andrews, St.~Andrews, UK}
\affil{Center~for~Computational~Astrophysics, Flatiron~Institute, New~York, NY, USA}
\author[0000-0002-0296-3826]{Rodrigo Luger}
\author[0000-0002-9328-5652]{Daniel Foreman-Mackey}
\affil{Center~for~Computational~Astrophysics, Flatiron~Institute, New~York, NY, USA}
\author[0000-0003-4859-2060]{Robert R. Howell}
\affil{Department~of~Geology~\&~Geophysics, University~of~Wyoming, Laramie, WY, USA}
\author[0000-0001-7619-652X]{Julie A. Rathbun}
\affil{Planetary~Science~Institute, Tuscon, AZ, USA}

\begin{abstract}
Jupiter's moon Io is the most volcanically active body in the Solar System with hundreds of active volcanoes varying in intensity on different timescales.
Io has been observed during occultations by other Galilean moons and Jupiter since the 1980s, using high-cadence near infrared photometry. 
These observations encode a wealth of information about the volcanic features on its surface.
    We built a generative model for the observed occultations using the code \textsf{starry} which enables fast, analytic, and differentiable computation of occultation light curves in emitted and reflected light.
    Our probabilistic Bayesian model is able to recover known hotspots on the surface of Io using only two light curves and without any assumptions on the locations, shapes or the number of spots.
    The methods we have developed are also directly applicable to the problem of mapping the surfaces of stars and exoplanets. \href{https://github.com/fbartolic/volcano}{\color{linkcolor}\faGithub}
\end{abstract}

\section{Introduction}
The surface of Io is covered with hundreds of volcanoes which appear as bright spots in the near infrared, their intensities varying on timescales ranging from days to decades.
The global heat flow on its surface is about 40 times larger than Earth's \citep{breuer2007,davies2010} and about 50\% of the heat flow emanates from only 1.2\% of Io's surface \citep{veeder2012}.
This intense volcanic activity cannot be explained by radioactive decay, the mechanism which drives Earth's volcanism.
Instead, the volcanism on Io is driven by tidal interactions with Jupiter and sustained by the Laplace resonance with Europa and Ganymede \citep{peale1979}.
Contrary to Earth, Io lacks plate tectonics.
The heat transport mechanism most likely operating on Io is volcanism generated in the lithosphere (the outermost shell of a terrestrial body).
This process is also known as \emph{heat-pipe volcanism}, whereby magma is transported to the surface through localised vents \citep{oreilly1981}.
Heat-pipe volcanism likely occurred in the early history of terrestrial planets in the Solar System \citep{moore2013,moore2017}, most notably on early Earth prior to the onset of plate tectonics.

Besides providing a window into early volcanic activity in the Solar System, Io is also in many ways an analogue of a volcanically active exoplanet.
Volcanic exoplanets, sometimes dubbed ``super-Ios'' or ``lava worlds'', have gathered a lot of interest in recent years.
Photometric and spectral signatures of volcanic activity on such worlds will likely be detectable in the near future with telescopes such as JWST and LUVOIR \citep{kaltenegger2010,henning2018,oza2019,chao2020}.
Existing exoplanet detections of planets with potential volcanic activity include CoRoT-7b \citep{barnes2010}, the first rocky exoplanet discovered and one which is likely heated by strong tidal forces; 55 Cancri e, whose inferred longitudinal offset in peak surface emission has been attributed to (among other things) lava flows on the surface \citep{demory2016,demory2016d,hammond2017a}; several planets in the TRAPPIST-1 system in a Laplace-like resonance likely exhibiting volcanic activity \citep{kislyakova2017,dobos2019}, and many others.
Not all such exoplanets are expected to have Io-like volcanism.
Some will have magma oceans because of their proximity to the star and others will have volcanism powered by nuclear decay, similar to volcanism on present day Earth.

Io has been observed extensively using both space and ground observatories.
High resolution images of Io's surface were taken by space missions such as Voyager \citep{smith1979}, Galileo \citep{belton1996} and Juno \citep{mura2020}.
The surface has also been resolved from ground based observations in the near infrared using disk resolved imaging \citep{howell1985,simonelli1986,spencer1990} and adaptive optics observations \citep{marchis2000,marchis2005,dekleer2016}.
Most importantly for this work,  starting with \cite{spencer1990} Io has been sporadically observed over a timespan of decades using high cadence near infrared photometry taken during occultations by Jupiter.
Occultations by Jupiter occur twice every orbit of ~1.7 days whereas occultations by Europa, Ganymede or Callisto (so-called ``mutual occultations'') happen every ~6 years when Earth passes through the orbital plane of the Galilean satellites.
Multiple occultations are then observable over a course of approximately one year.
The majority of occultations by Jupiter are observed when Io is in Jupiter's shadow (``in eclipse'') while mutual occultations are almost always observed in sunlight.
Only the brightest volcanoes are visible over the reflected light background when Io is illuminated by the Sun \citep{veeder1994,dekleer2016a}.

Both kinds of occultations have been used to study volcanic activity on the surface.
\cite{spencer1994} observed several occultations of Io by Europa and detected a major brightening of the most powerful of Io's volcanoes, Loki, relative to previous Voyager observations.
\cite{rathbun2002},\citet{rathbun2006} and \citet{rathbun2010} used observations from NASA's 
Infrared Telescope Facility (IRTF) telescope to study the long term variability of different volcanic spots, finding evidence of periodicity in Loki's eruptions and establishing the transient nature of observed emission from most volcanoes. 
By studying a (spatially resolved) occultation of Io by Europa, \cite{dekleer2017} mapped the Loki Patera (\emph{Patera} is a type of an irregular crater) region to a precision of about 2 kilometers.
In addition to observations of occultations in the near infrared, several groups have been observing mutual occultations in the optical for decades with the purpose of inferring the optical albedo of Io and improving ephemeris precision for Galilean satellites \citep[][and references therein]{arlot1974,saquet2018,morgado2016a}.
Understanding the detailed albedo distribution is crucial to constraining the ephemeris of Io to a very high precision.

Most studies of (unresolved) occultations of Io fit multiple light curves independently with the goal of inferring one-dimensional longitudinal variations in brightness on its surface (notable exceptions are \cite{spencer1994} and \cite{dekleer2017}), assuming relatively strong priors on the locations of individual volcanoes.
In this work, we build a fully probabilistic model in order to infer a \emph{two-dimensional map} of Io's volcanic surface from archival IRTF observations of Io in the near infrared.
We assume that the map corresponding to a given set of light curves is static (up to an overall amplitude), that is, that it didn't change substantially between the times of observations.
In Paper II in this series we will relax this assumption by constructing a model which can be used to infer a time-variable map of Io from an any number of light curves observed at arbitrary times.

The model relies on the recently developed code \textsf{starry}\footnote{\url{https://starry.readthedocs.io/en/latest/}} \citep[][Luger et al. 2021 in prep]{luger2019a} which enables fast analytic computation of occultation light curves and phase curves for objects whose surface can represented in terms of spherical harmonics.
\textsf{starry} can compute phase curves and occultations in both emitted light (for modeling the isotropic thermal emission from Io's surface) and reflected light (for mapping albedo variations).
It is many orders of magnitude faster and more accurate than pixel based algorithms and it computes exact gradients of the flux with respect to all parameters through automatic differentiation.
\textsf{starry} also comes with extensive tools for visualizing spherical harmonic maps and simulating data.

Our main goal in this paper is to develop a general model for mapping surfaces of occulted bodies with sparse, high contrast features and apply it to observations of Io.
Because of high resolution resolved imaging of Io, we can compare features in our inferred maps to known volcanic hotspots and test the sensitivity of the inferred maps to different choices of priors and noise models.
The paper is organized as follows.
In \S\ref{sec:data} we describe the IRTF light curves of occultations by Jupiter which we use to infer maps.
In \S\ref{sec:model} we discuss the generative model for the data and we write down the likelihood function.
In \S\ref{sec:inverse_problem} we focus on the question of how to do Bayesian inference given the forward model specifications defined in \S\ref{sec:model}.
We discuss and test different priors on map features, we quantify the information content of light curves and fit the models using Hamiltonian Monte Carlo (HMC). 
We then test the model on realistic simulated data.
In \S\ref{sec:results} we show the results on real IRTF data for two pairs of light curves, one pair observed in 1998 and another in 2017.
We show the inferred maps and the parameters quantifying the location and intensity of the hot spots, we discuss the sensitivity of the results to different choices of priors and plot the inferred hotspots on an optical map of Io constructed from Galieleo observations.
Finally, in \S\ref{sec:discussion} we summarize the main results of the paper and discuss potential applications of our model to observations of volcanic exoplanets.

\vspace{1em}

The code associated with this paper is publicly hosted in a GitHub repository\footnote{\url{https://github.com/fbartolic/volcano}}. 
All of the figures in this paper were auto-generated
using the Azure Pipelines continuous integration (CI) service.
Icons next to each of the figures \codeicon \,
link to the exact script used to generate them to ensure the reproducibility
of our results. 

\section{Data}
\label{sec:data}

\begin{figure}[t!]
    \begin{centering}
    \includegraphics[width=\linewidth]{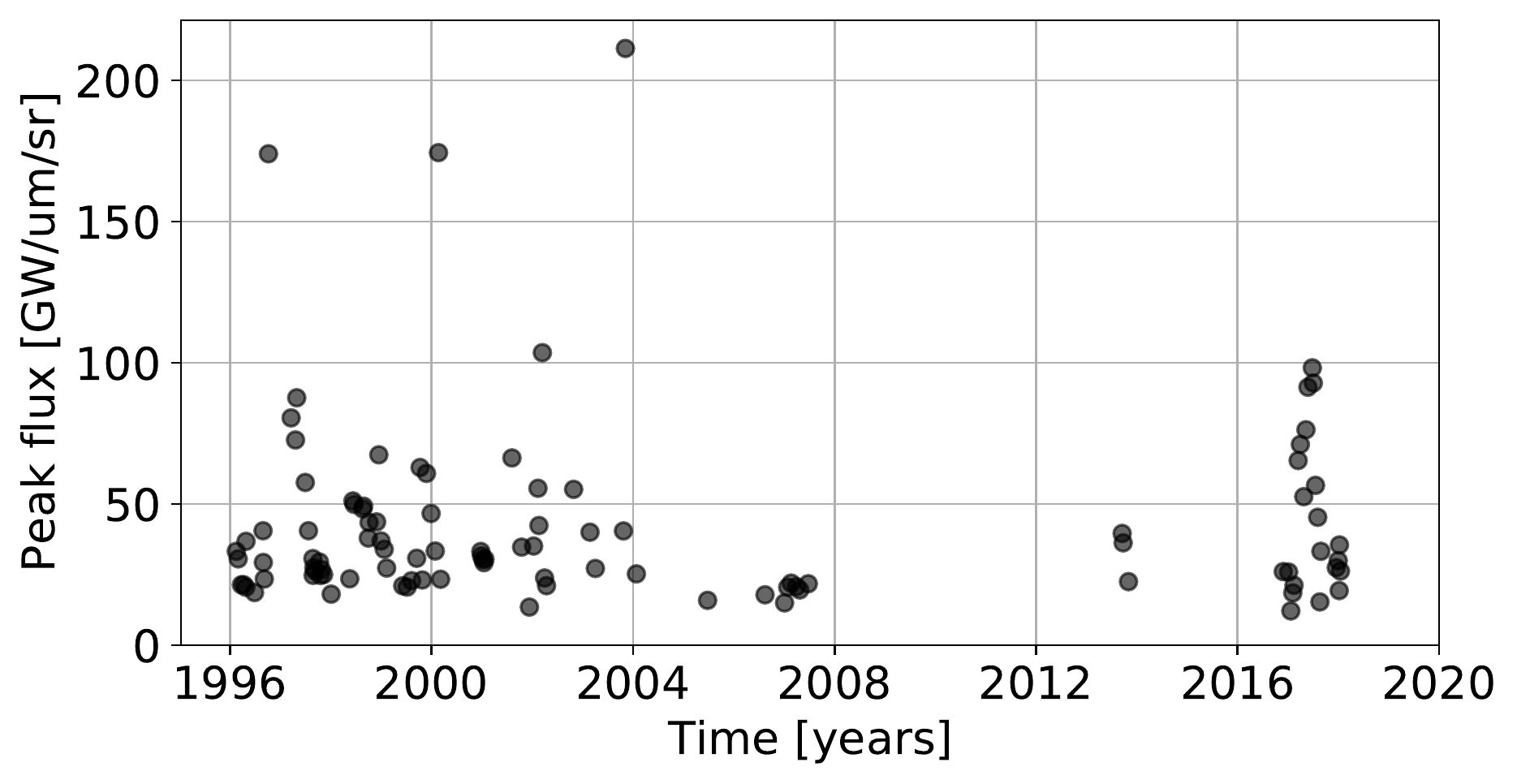}
    \oscaption{irtf_dataset}{%
        Maximum flux for each IRTF light curve of Io observed during an occultation of Io by Jupiter.
       \label{fig:irtf_max_flux}
    }
    \end{centering}
\end{figure}

The complete dataset consists of 112 sets of observations of Io in the near infrared taken from 1996 until 2018 using different instruments at NASA's IRTF observatory, the NSFCam  1-5 $\mu m$ camera \citep{shure1994}, the SpeX 0.8-5.5 $\mu m$ spectrographer/imager \citep{rayner2003} and the iShell 1.1-5.3 $\mu m$  spectrographer/imager \citep{rayner2016}.
Although the observations span decades, the timespan between observations varies and there were very few observations taken between 2008 and 2016.
All of the observations were taken while Io was \emph{in eclipse}, meaning that it was in Jupiter's shadow so the source of observed emission from the surface is thermal radiation.
On the other hand, for observations of Io \emph{in sunlight} the observed flux is a sum of thermal (volcanic) emission and reflected sunlight which depends on the albedo map in the near infrared.
In figure~\ref{fig:irtf_max_flux} we plot the maximum flux for each occultation light curve in units of $\mathrm{GW}/\mu \mathrm{m}/\mathrm{sr}$ as a function of time.
This value is approximately equal to the total flux emitted from the Jupiter-facing side of Io.
The major variation in the baseline brightness is driven by Loki.
Most hotspots on Io can be divided into those who are persistently active for a year or longer at moderate intensity, and those which have bursts of activity and reach very high intensities for short periods of time \citep{dekleer2016}.

\begin{figure}[h!]
    \begin{centering}
        \includegraphics[width=\linewidth]{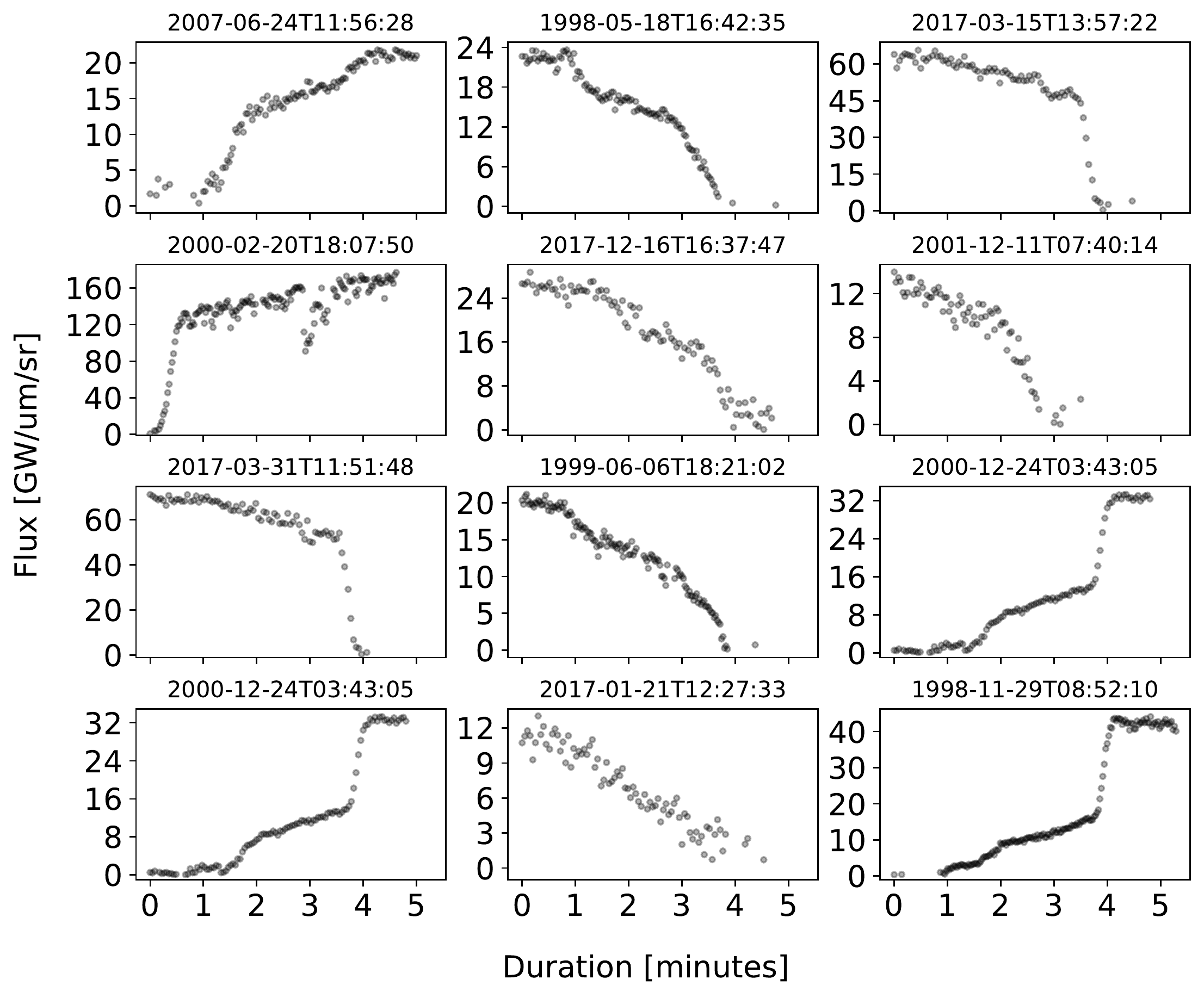}
    \oscaption{irtf_dataset}{%
        A selection of sample light curves observed using NASA's IRTF telescope during occultations of Io by Jupiter in our dataset.
        The step-like morphology of the light curves is due to bright volcanoes on Io's surface coming in or out of view during the course of an occultation.
        These light curves visibly encode information about the features on the surface.
        \label{fig:irtf_sample_lightcurves}
    }
    \end{centering}
\end{figure}

In Figure~\ref{fig:irtf_sample_lightcurves} we display a random subset of all the light curves
in the dataset. 
Each light curve covers a timespan of $\sim 4$ minutes which is the duration of the ingress (or egress) part of the entire occultation.
The exposure cadence for each light curve is about a second, with some variation between light curves.
It follows that over the course of a single exposure Jupiter's limb passes over about 15km on Io's surface which provides a lower bound for the size of the features that we can reliably estimate.
The shapes of light curves strongly deviate from a smooth variation that one would expect to see if the surface thermal emission was uniformly distributed.
Especially prominent are light curves with clear step-like features which are due to bright localized spots coming in and out of view during the course of an occultation.
The fact that these features are so clearly visible means that even individual light curves encode a wealth of information about the spots.

The photometric quality of the light curves varies from year to year because multiple instruments were used over the years. 
As with all ground-based photometry, the observations are influenced by atmospheric variability which results in some correlated noise in the light curves, because of this the flux is not always monotonically increasing or decreasing as one would expect.
Complicating matters, the light curves lack errorbars. 
We deal with this by treating all errobars as free parameters with a common characteristic scale in the final model.

\section{Model}
\label{sec:model}
In this section we describe the generative model for the occultation light curves of Io.
To simulate the observed light curves, we first need to specify the geometry of each occultation event (Section~\ref{ssec:orbital_parameters}), we then need to specify what the surface of Io looks like in a spherical harmonic basis and compute the theoretical flux at the times of observations using \textsf{starry} (Section~\ref{ssec:model_spec}), finally, we need to specify a noise model for the observed data (Section~\ref{ssec:likelihood}).

\subsection{Orbital parameters}
\label{ssec:orbital_parameters}
To compute the geometry of every occultation, we use the \href{https://ssd.jpl.nasa.gov/horizons.cgi}{JPL Horizons database} which uses the latest ephemeris for the position of Jupiter \citep{folkner2014}, and its satellites \citep{jacobson2015}.
The ephemeris is usually accurate to a few kilometers on Io's surface.
We access JPL Horizons through the Python package \textsf{astroquery} \citep{ginsburg2019}.
To compute the relative position of Jupiter and Io we need the right ascension and declination and to to fix the orientation of Io we need the longitude and the latitude at the center of Io's disc as seen from Earth (\textsf{Ob-lon} and \textsf{Ob-lat} in \textsf{astroquery}) and the counterclockwise angle between the celestial north pole unit vector projected onto the plane of the sky and Io's north pole (\textsf{NP.ang}).
All longitudes provided by Horizons are positive  in the direction of west.
Horizons provides all ephemeris with a minimum cadence of 1 minute, we interpolate these values so that we can evaluate them for arbitrary times.

The coordinate system in \textsf{starry} is defined to be right handed such that the $\hat{z}$
axis points towards the observer and the $\hat{x}$ axis points to the right on the plane of the sky.
The radius of the occulted sphere is fixed to 1 and the orientation is specified by three angles,
the counterclockwise obliquity angle \textsf{obl} between the $\hat{y}$ axis and the north pole of the sphere, the inclination angle \textsf{inc} which is set to $90^\circ$ if the north pole is aligned with the $\hat{y}$ axis, and the phase angle \textsf{theta} which rotates the sphere around the $\hat{y}$ axis in the eastward direction.
Expressed in terms of Horizons variables, these angles are given by $\mathrm{obl}=\textsf{NP.ang}$, $\mathrm{inc}=90^\circ-\textsf{Ob-lat}$ and $\mathrm{theta}=\textsf{Ob-lon}$.
The occultor position relative to the occulted object is given by
\begin{align}
    \mathrm{xo}/\gamma&=-\Delta\alpha\,\cos\delta\\
    \mathrm{yo}/\gamma&=\Delta\delta\\
    \mathrm{zo}/\gamma&=1
\end{align}
where $\Delta\alpha$ and $\Delta\delta$ are differences in right ascension and declination respectively relative to the occulted object and $\gamma$ is the angular radius of the occulted sphere.

\subsection{Jupiter's effective radius}
\label{ssec:effective_radius}
Although the sky position of Jupiter relative to Io is known to a precision of a few kilometers, several complications arise when attempting to compute Jupiter's radius.
First, Jupiter is not spherical and its equatorial radius is greater than the polar radius by thousands of kilometers. 
Because \textsf{starry} does not (yet) support occultations for non-spherical objects and because the effect is negligible at the resolution of our maps, we assume that Jupiter is locally spherical at the point of an occultation. 
We estimate an effective radius from the measurements of Jupiter's shape from Voyager radio occultation data \citep[Fig.~7]{lindal1981}.
When computing the occultation latitude we fix Jupiter's latitude to the value at the center of Io's disc because 
the variation of Jupiter's effective radius along Io's disc is negligible.

Second, Jupiter is gaseous so it does not have a well defined boundary.
In principle, we should compute an effective radius of Jupiter at different altitudes (pressure) in the atmosphere and model an occultation of Io by a fuzzy occultor.
Although this is possible with \textsf{starry}, it is unnecessary for our models because the characteristic scale height of Jupiter is around 27 km  which is below the uncertainty of our inferred maps.
We instead follow the approach in \cite{spencer1990} and compute the effective radius of Jupiter at about 2.2 mbar, the pressure (and the associated effective radius) at which a  bright source on the surface of Io fades by 50\% due to differential refraction during the course of an occultation.

%These data come in the form of a plot of effective radius  as a function of Jupiter's planetocentric latitude at a fixed pressure of 100 mbar \citep[Fig.~7 in ][]{lindal1981}.
%

Third, the information on the effective radius in \cite{lindal1981} is provided only at a fixed pressure of 100 mbar.
To adjust the values for a lower pressure of 2.2 mbar we assume an exponential pressure profile $P=P_0\, e^{-\Delta r/H}$ where $H$ is the scale height and $\Delta r$ is the height difference between the two pressure levels.
It follows that to convert the shape profile at 100 mbar to 2.2 mbar we need to add the factor
$-H\ln(2.2/100)$ which is assumed to be constant in the $\pm 21$ deg latitude range in which the occultations occur.
In addition to refractive absorption in Jupiter's atmosphere there is a also slight additional molecular absorption due to methane which \cite{spencer1990} estimate to be equal to around 12\% in their filter, we choose to ignore this because it is far below the resolution of our maps.

Finally, we have to account for the fact that the light from Io is getting significantly bent at the point of half reflective intensity in Jupiter's atmosphere which results in a smaller projected limb of Jupiter on the surface of Io then would be the case for straight propagation.
It is zero at the beginning of the disappearance of a hot spot when there is no refraction, increasing to one scale height $H$ at the half intensity point and then increasing further to a large value when Io disappears behind Jupiter's limb.
We ignore the variation in the bending and adopt a fixed value of one scale height for this effect which we subtract from the value of the effective radius.

In summary, to compute an effective radius of Jupiter for a given occultation of Io first we have to compute Jupiter's planetocentric latitude at which Io's disk disappears or reappears behind the limb, then use a modified shape profile from \cite{lindal1981} to get an effective radius and finally we have to subtract one scale height due to light bending.
There are substantial uncertainties in each of these steps.
The shape profile data are quite old and it is not known if the structure of Jupiter's has remained constant since the 1980s. 
The shape profile also depends on the wind velocity structure and temperature which we do not take into account.
In addition to uncertainties about the atmospheric structure, there are uncertainties associated with digitizing the data shown in Fig.~7 in \cite{lindal1981} because it is not available in table form.
Since our main focus in this work is the map model, we leave a detailed investigation into the various sources of error that go into the radius estimate for future work.

\subsection{Linear model for the flux}
\label{ssec:model_spec}
Given the geometry of on occultation event at the times of observations, computing the predicted flux with \textsf{starry} is straightforward.
\textsf{starry} computes the integrated flux of an unocculted or occulted sphere analytically given an expansion of the surface map in spherical harmonics $Y_{lm}(\theta,\phi)$ up to a certain \emph{degree} $l$.
The map is defined by a vector of spherical harmonic coefficients $\mathbf{y}$ which is dotted into the spherical harmonic basis vector $\left(Y_{0,0}\;Y_{1,-1}\;Y_{1,0}\;Y_{1,1}\;Y_{2,-2}\;Y_{2,-1}\;Y_{2,0}\;Y_{2,1}\;Y_{2,2}\; \cdots\right)^{\top}$.
The total number of spherical harmonic coefficients is $(l+1)^2$.
In addition to computing thermal phase curves and occultation light curves, \textsf{starry} also solves the considerably more complex problem of computing reflected light phase curves and occultations in which the occulted object is illuminated by a distant light source \citep[Luger et al. 2021 in prep][]{}. 
In that case the coefficient vector $\mathbf{y}$ represents spherical albedo.
For observations of Io in sunlight in a wavelength range where where both the thermal component and the reflected sunlight component of observed flux are comparable, it would be necessary to simultaneously fit for two vectors of spherical harmonic coefficients, one for the albedo distribution and the other for thermal emission.
In this paper we only use observations of Io in eclipse so we do not need to model the reflected light component.

Conditioned on fixed parameters specifying the geometry of the occultations, the \textsf{starry} model is \emph{linear} for both emitted and reflected light maps.
\cite{luger2019} achieved this by representing all rotations, changes of basis transformations and integrals with complicated boundaries which are needed to compute the flux as linear transformations.
Since a sequence of linear mappings is also linear, the predicted flux can be written as
\begin{equation}
    \mathbf{f}=\mathbf{A}\,\mathbf{y}
    \quad,
    \label{eq:linear_model}
\end{equation}
where the column vector $\mathbf{f}$ of shape $(T, 1)$ is the predicted flux for different values of the occultor position and $\mathbf{A}$ is the design matrix \citep[see appendix B.1. in ][]{luger2021a} of shape $(T, N)$ (with $N=(l+1)^2$) which encodes all operations needed to compute the integrated flux at different viewing angles of the occulted sphere.
If the geometry isn't known precisely the matrix $\mathbf{A}$ is not fixed and the model is in principle no longer linear although one can still use the fact that it is linear when conditioned on a particular values of the nonlinear parameters to speed up inference.

The characteristic angular size of features that can be represented with a given map is set by the degree of the map and it is approximately equal to $180^\circ/l$.
For reference, state of the art inferences involving phase curves and secondary eclipses of exoplanets are able to constrain features of order $l=1$ (inferring a bright spot offset from the substellar point) but for Io we need to fit much higher order maps because the typical scale of volcanic spots is on the order of tens of kilometers (a few degrees).
\textsf{starry} can handle occultations up to $l\approx 20$ before numerical instabilities kick in \citep{luger2019a} which corresponds to a minimum resolution of $9^\circ$ which means that we are not able to constrain the physical size of the spot.
We discuss the implication of this resolution limit in Section~\ref{sec:results} and Section~\ref{sec:discussion}.

Although \textsf{starry} was built around the idea of expanding surface features in a spherical harmonic basis in which we can compute all fluxes analytically, this basis may not be ideal for doing inference because it can be difficult to encode assumptions (priors) on what we expect the map to look like.
For example, the most important constraint on the map we would like to incorporate in the model is that the intensity of the map is positive at every location. 
This constraint is important not only because we want to avoid having unphysical regions in inferred maps but also because it imposes a very strong prior on the map which substantially reduces the difficulty of inference when the data is not particularly informative.

There are two ways of enforcing positivity that we are aware of, both of which involve evaluating the intensity on a fixed grid of \emph{pixels} and disallowing negative values of those pixels.
The first is to fit for the spherical harmonic coefficients $\mathbf{y}$, evaluate the pixel grid at each MCMC step and reject all samples of the coefficient vector $\mathbf{y}$ which result in pixels with negative intensity.
The issue with this procedure is that rejecting samples in this way implies a prior probability distribution on $\mathbf{y}$ which cannot be mapped to a parameter space with infinite support and Hamiltonian Monte Carlo does not work well with constrained parameter spaces.
The second approach is to dispense with the spherical harmonics entirely and compute the full model using a high resolution grid of pixels.
In practice this is very difficult to do because we would need a very high resolution grid to compute the light curves accurately with minimal discretization noise.  
\textsf{starry} uses spherical harmonics as a basis precisely to avoid this problem.
\textbf{Instead, we opt for a hybrid approach in which we fit for pixels but at each MCMC step we convert the pixels to spherical harmonic coefficients to compute the light curve.}

To implement the hybrid approach we need to be able to switch between spherical harmonics and pixels.
Transforming spherical harmonics $\mathbf{y}$ to pixels $\mathbf{p}$ is straightforward because there exists a linear operator $\mathbf{P}$ which maps $\mathbf{y}$ to $\mathbf{p}$:
\begin{equation}
    \mathbf{p}=\mathbf{P}\,\mathbf{y}
    \label{eq:ylms_to_pixels}
\end{equation}
Each row of $\mathbf{P}$ contains values of each of the spherical harmonic coefficients at a given point on the grid. 
To construct the grid we use an equal area Molleweide projection in order to have a uniform distribution of pixels across the sphere.
The grid needs to be fine enough to ensure that the intensity is positive over most of the sphere.

Switching from pixels to spherical harmonics is somewhat more complicated because $\mathbf{P}$ is in general not a square matrix so we cannot compute its inverse to obtain the inverse transform.
Instead, we can compute an approximate inverse (a pseudoinverse) $\mathbf{P}^\dagger$ by solving the linear system $\mathbf{P}\,\mathbf{P}^\dagger=\mathbf{I}$, where $\mathbf{I}$ is the identity matrix. 
The solution is given by
\begin{equation}
\mathbf{P}^\dagger=\left(\mathbf{P}^{\top} \mathbf{P}+\lambda \mathbf{I}\right)^{-1} \mathbf{P}^{\top}
    \quad,
\end{equation}
where $\lambda$ is a small regularization parameter and $\mathbf{I}$ is the identity matrix.
The mapping from pixels to spherical harmonics is then given by
\begin{equation}
    \mathbf{y}\simeq\mathbf{P}^\dagger\mathbf{p}
    \label{eq:pixels_to_ylms}
\end{equation}
Both $\mathbf{P}$ and $\mathbf{P}^\dagger$ can be precomputed to speed up inference.
When using pixels to impose a positivity constraint on the spherical harmonic map we need to make sure that the number of pixels is greater than the number of spherical harmonic coefficients by a factor of a few to ensure positivity approximately everywhere on the sphere.
In practice we find that we need to use at least 4 times as many pixels as spherical harmonics which means that the computational cost of this model is higher than if we just fit for spherical harmonics. 

\begin{figure}[t!]
    \begin{centering}
    \includegraphics[width=1.\linewidth]{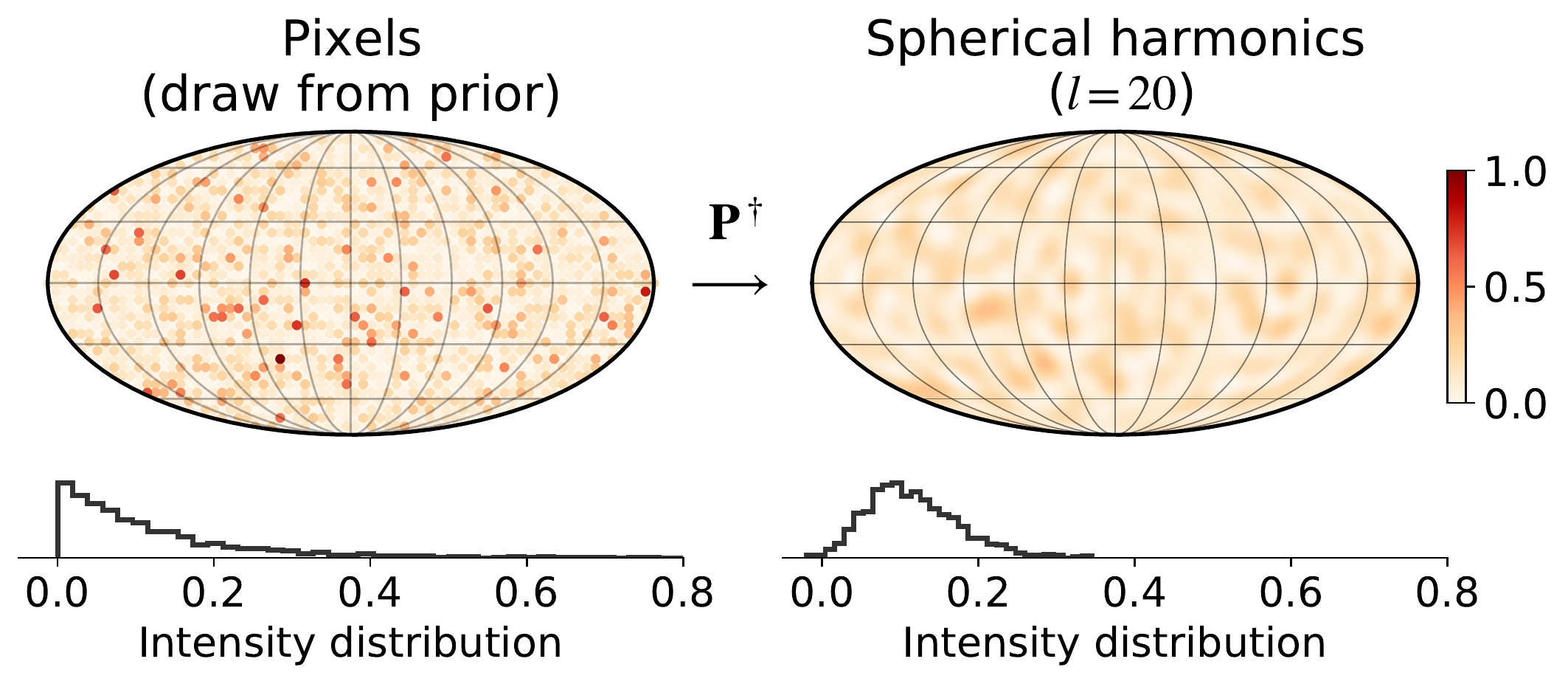}
    \oscaption{pixels_to_spherical_harmonics}{%
        Illustration of the transformation from a pixel map (left) defined on a fixed grid on the sphere into the spherical harmonic basis at $l=20$ (right) via a linear operator $\mathbf{P}^\dagger$.
        The intensity of each pixel was drawn independently from an exponential prior.
        The histograms below each of the maps show the distributions of intensities on the map. 
        Since the pixel map is higher resolution than the spherical harmonic map at $l=20$, the         spherical harmonic map appears smoother and the intensity distribution is more similar to a skewed Gaussian rather than an exponential distribution.
        Despite this fact, we find that fitting for maps in the pixel basis and transforming them to spherical harmonics at each MCMC step in order to compute the light curve analytically with \textsf{starry} is better than fitting for spherical harmonics because the pixel grid provides a strong constraint on the spherical harmonic map structure.
  \label{fig:pixels_to_harmonics}}
    \end{centering}
\end{figure}

To summarize, in our hybrid model we first construct a fixed high resolution pixel grid in latitude and longitude, then use Equation~(\ref{eq:pixels_to_ylms}) to convert pixels to spherical harmonics and finally use Equation~(\ref{eq:linear_model}) to evaluate the model.
Although we fit for the pixels $\mathbf{p}$ we store the spherical harmonic coefficient vectors $\mathbf{y}$ as the final product of the inference. 
Figure~\ref{fig:pixels_to_harmonics} illustrates the transformation from the pixel basis to the spherical harmonic basis via $\mathbf{P}^\dagger$.
On the left we show the pixel map where each pixel was independently drawn from an exponential prior. 
On the right is the same pixel map transformed to a spherical harmonic basis via $\mathbf{P}^\dagger$. 
The histograms underneath each map show the distribution of intensities.
Since the pixel map is higher resolution than the spherical harmonic map it can only be approximately represented at a finite order of the spherical harmonic expansion so the map on the right appears to be smoother and the intensity distribution is more similar to a skewed Gaussian with a heavy right tail than an exponential distribution.
Nevertheless, we find that setting priors on pixels is a far better solution than fitting the spherical harmonic coefficients directly and is worth the extra computational cost which comes with the increased dimensionality of the parameter space (see Section~\ref{ssec:pixels_vs_harmonics} for a demonstration).

\subsection{The likelihood}
\label{ssec:likelihood}
Finally, we have to specify the noise model which means we have to define a likelihood function.
Assuming we have a single light curve with $T$ data points and a map defined by the pixels $\mathbf{p}$, the (Gaussian) log likelihood is given by
\begin{equation}
    \ln\mathcal{L}=-\frac{1}{2}\left[\mathbf{f}_\mathrm{obs}-\mathbf{f} \right]^{\top}
    \boldsymbol{\Sigma}^{-1}\left[\mathbf{f}_\mathrm{obs}-\mathbf{f} \right]
    \quad,
    \label{eq:likelihood}
\end{equation}
where $\mathbf{f}_\mathrm{obs}$ is the observed light curve, $\boldsymbol{\Sigma}$ is the data covariance matrix and $\mathbf{f}$ is predicted flux given by
\begin{equation}
    \mathbf{f}=\mathbf{A}'\,\mathbf{p} +b\,\mathbf{1}_T
    \quad,
    \label{eq:flux_model}
\end{equation}
where $\mathbf{A}'\equiv\mathbf{A}\mathbf{P}^\dagger$ and $b$ is a constant flux offset parameter which we have added to account for stray flux which cannot be attributed to Io.
Depending on the observation, this flux is most often residual light from Jupiter.

To model the data covariance $\boldsymbol{\Sigma}$ we use use a Gaussian Process and we compute the likelihood using the fast Celerite method \citep{foreman-mackey2017} as implemented in the \textsf{celerite2} package \citep{foreman-mackey2017a,foreman-mackey2018}.
We use the simple (approximate) Mat\'ern 3/2 kernel function which is parametrized by two values, a standard deviation parameter $\sigma_\mathrm{GP}$ and a characteristic timescale parameter $\rho_\mathrm{GP}$.
The Mat\'ern 3/2 kernel is defined by
\begin{equation}
    k(\tau;\sigma_\mathrm{GP},
    \rho_\mathrm{GP})=\
    \sigma_\mathrm{GP}^{2}\left(1+\frac{\sqrt{3} \tau}{\rho_\mathrm{GP}}\right) \exp \left(-\frac{\sqrt{3} \tau}{\rho_\mathrm{GP}}\right)
    \quad,
\end{equation}
where $\tau=|t_n-t_m|$ and $\epsilon$ controls the quality of the approximation.
A single element of the data covariance matrix $\boldsymbol{\Sigma}$ is then 
\begin{equation}
    \boldsymbol{\Sigma}_{nm}=\sigma_n^2\delta_{nm} + k(\tau;\sigma_\mathrm{GP},\rho_\mathrm{GP})
    \quad,
    \label{eq:data_covariance_element}
\end{equation}
where $\sigma_n$ is the errorbar for the n-th data point.
Since we fit multiple independent light curves the total log likelihood is the sum of individual likelihoods defined in Equation~(\ref{eq:likelihood}).

\section{The inverse problem}
\label{sec:inverse_problem}
Having defined a probabilistic model which describes how to compute a realistic light curve for an occultation of Io in the previous section, in this section we discuss the inverse problem of inferring a surface map by fitting a set of occultation light curves in a Bayesian framework.

\subsection{The information content of a light curve}
\label{ssec:information_content}
The mapping problem is famously ill posed, meaning that specific linear combinations of spherical harmonic coefficients will be in the nullspace of the linear mapping  $\mathbf{A}$ in Equation~(\ref{eq:linear_model}) \citep{luger2021}.
This means that in general, even if we had noiseless observations, it would still be impossible to recover certain features on the surface.
To recover the greatest information about the surface we need to have a mechanism which breaks the various degeneracies.
For example, with phase curves we can recover primarily longitudinal variations in emission.
Occultations are substantially better because the limb of the occultor sweeps across the surface of the occulted sphere, thereby exposing or blocking light from different points on the surface.
An ideal set of observations would consist of phase curves together with observations of multiple occultations by a small occultor at different latitudes and different phases.
Phase curves and occultations in reflected light are even more informative because of the nonuniform illumination profile of the incident radiation and the presence of a day/night terminator line \citep[][Luger et al. 2021 in prep]{luger2019a}.
In some cases for reflected light observations (phase curves of an inclined planet for example) there can even be no nullspace \emph{at all} low spherical harmonic degrees.

We can reformulate these statements on how useful given observations are more precisely by computing a measure of their information content.
Given that our model is linear, assuming Gaussian priors on the spherical harmonic coefficients with covariance $\boldsymbol{\Lambda}_\mathbf{y}$ and a Gaussian likelihood, the posterior can be computed analytically and its mean is given by
\begin{equation}
    \widehat{\mathbf{y}}=\boldsymbol{\Sigma}_{\hat{\mathbf{y}}}\left(\mathbf{A}^{\top} \boldsymbol{\Sigma}_{\mathbf{f}}^{-1} \mathbf{f}+\boldsymbol{\Lambda}_{\mathbf{y}}^{-1} \boldsymbol{\mu}_{\mathbf{y}}\right)
    \quad,
    \label{eq:linear_solve_mean}
\end{equation}
where the posterior covariance matrix $\boldsymbol{\Sigma}_{\hat{\mathbf{y}}}$ is 
\begin{equation}
\boldsymbol{\Sigma}_{\hat{\mathbf{y}}}=\left( \mathbf{A}^{\top} \boldsymbol{\Sigma}_{\mathbf{f}}^{-1} \mathbf{A} +\boldsymbol{\Lambda}_{\mathbf{y}}^{-1}\right)^{-1}
    \label{eq:linear_solve_cov}
\end{equation}
We define the information content as the variance reduction of the posterior relative to the prior which is called \emph{posterior shrinkage}.
The posterior shrinkage $S$ is defined as \citep{luger2021a,betancourt2018}:
\begin{equation}
S \equiv 1-\lim _{\sigma_{0}^{2} \rightarrow \infty} \frac{\sigma^{2}}{\sigma_{0}^{2}}
    \quad,
\end{equation}
where $\sigma^2_0$ is the prior variance and $\sigma^2$ is the posterior variance for a given spherical harmonic coefficient.
It tells us how well we can constrain a particular coefficient in the limit of infinite SNR observations.
Posterior shrinkage of 1 indicates that the data provides perfect information on the parameters while 0 indicates no gain in information relative to the prior.

In order to compute the posterior shrinkage we first compute the design matrices $\mathbf{A}$ with \textsf{starry} for different kinds of observations of Io in the period starting on the 1st of January 2009 and ending on the 1st of May 2010.
We chose this period because it covers the full season of mutual occultations which occur every 6 years.
We compute $\mathbf{A}$ for all observable occultations of Io by Galilean moons during that period, for occultations of Io by Jupiter and for phase curve observations.
The purpose of this is to determine the upper bound on what we can learn about the surface. 
We take the ephemeris from \textsf{JPL Horizons} and assume that it is known exactly; we also assume all observations are observations of thermal emission independent of whether Io is in sunlight or in eclipse because we are interested in constraining the volcanic emission rather than the albedo.

\begin{figure}[t!]
    \begin{centering}
    \includegraphics[width=\linewidth]{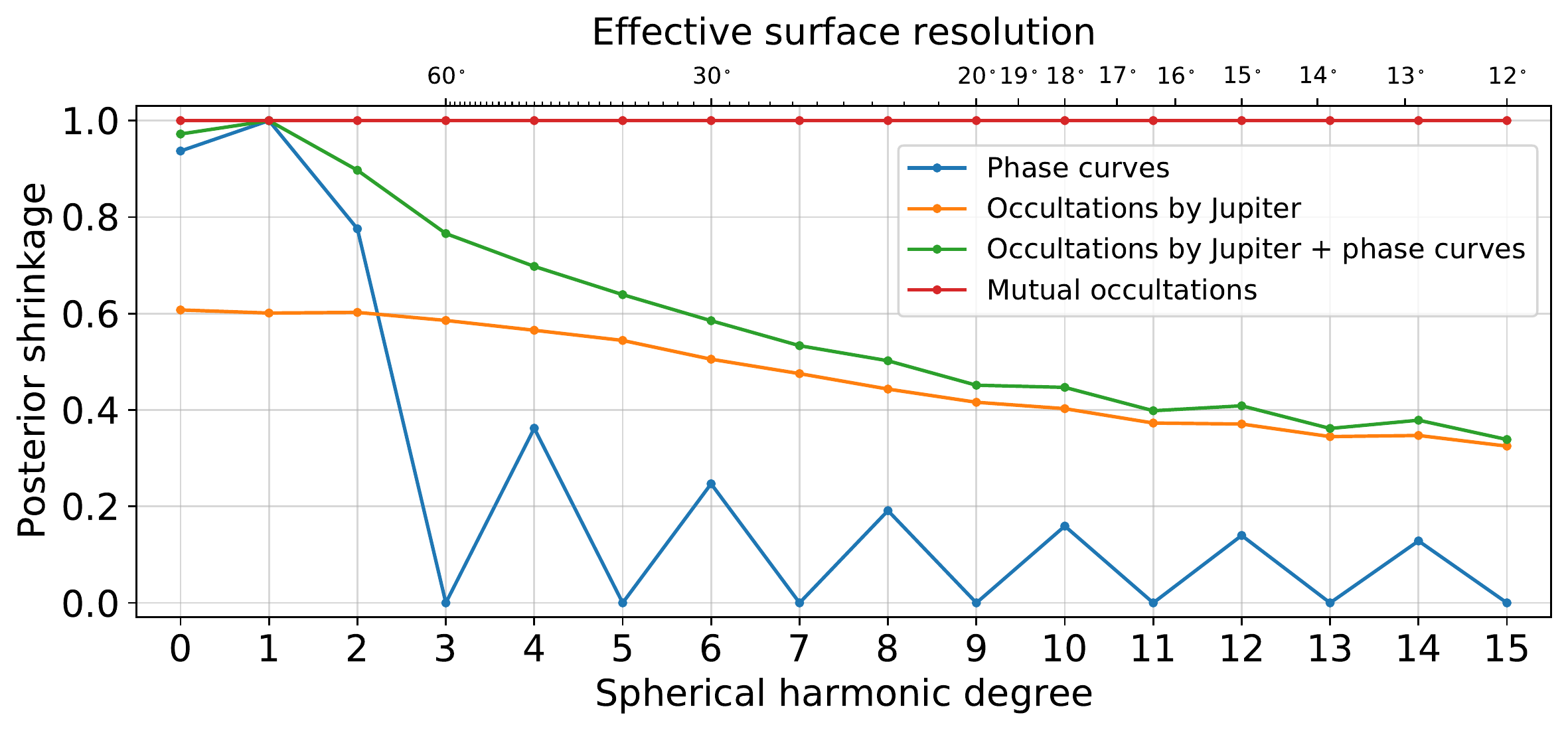}
    \oscaption{information_content}{
        The posterior shrinkage for different kinds of observations of Io as a function of spherical harmonic degree (angular scale), averaged across all $m$ modes.
    Posterior shrinkage of 1 represent maximum information gain in updating from the prior to the posterior while 0 represents no information gain. 
    The posterior variance has been computed for different kinds of simulated observations of Io over the course of a single year: phase curves (blue), occultations by Jupiter (orange), combined phase curves and occultations by Jupiter (green) and occulltations of Io by other Galilean moons (red lines). 
        \label{fig:information_content}
    }
    \end{centering}
\end{figure}

Fig.~\ref{fig:information_content} shows the posterior shrinkage as a function of $l$ (averaged over all $m$ modes) for phase curve observations (blue lines), occultations by Jupiter (orange), the former two combined (green), and mutual occultations by other Galilean moons (red).
As expected, the mutual occultations of Io by other Galilean moons are by far the most informative with posterior shrinkage of unity at all angular scales considered. 
Occultations by Jupiter are less informative because we only see one side of Io during an occultation.
Shrinkage for phase curves at odd degrees above $l=2$ is exactly zero because these coefficents are in the nullspace for objects rotating about an axis perpendicular to the line of sight and therefore cannot be constrained using only phase curves \citep{luger2021}. 
Although observations of mutual occultations most easily break the degeneracies, the drawback of these types of observations is that they only happen every 6 years and they almost never happen while Io is in eclipse, meaning that only the brightest volcanoes are visible above the reflected sunlight.
This fact is not captured in Figure~\ref{fig:information_content}. 

Figure~\ref{fig:information_content} gives us some idea about which kinds of observations are most informative but it doesn't really tell us how well we can constrain bright spot-like features we expect to see on Io.
To answer this question we have to create a simulated dataset and conduct the whole inference process.

\subsection{Pixels vs. spherical harmonics}
\label{ssec:pixels_vs_harmonics}
We use \textsf{starry} to generate a single simulated light curve of an occultation of Io by Jupiter from an $l=30$ map with known coefficients.
The simulated map consists of a spherical harmonic expansion of a bright spot with a Gaussian profile which we add to a uniform brightness map using the built in \textsf{add\_spot} function in \textsf{starry}.
The expansion is in the quantity $\cos(\Delta\theta)$ where $\Delta\theta$ is the angular separation between the center of the spot and another point on the surface of the sphere. 
We place the spot at $13^\circ$ N latitude and $51^\circ$ E longitude, we set the diameter of the spot ($2\Delta\theta$) to $5^\circ$ and we set the amplitude of the spot such that the total \emph{luminosity} of the map increases by 50\% with the addition of the spot.
We generate two light curves with 150 data points each, one for the duration of the ingress of the full occultation and the other for the egress.
Because the limb of the occultor sweeps over the disc of Io at different angles during ingress and ingress, this makes it possible break most degeneracies in the map and recover the location of the simulated spot.
We set the phase of the simulated map to be $10^\circ$ E at the beginning of ingress and $10^\circ$ W at the end of egress.
We assume that the geometry of the occultation is known exactly and we set the errorbars such that SNR=50 where the signal is defined to be the maximum value of the computed flux.
    We use this dataset to test the difference between setting a prior in the spherical harmonic basis $\mathbf{y}$ and in the pixel basis $\mathbf{p}$ by fitting an $l=20$ map to the dataset.

In the spherical harmonic model we place a Gaussian prior on $\mathbf{y}$ with covariance $\boldsymbol{\Lambda}=\mathrm{diag}(1^2,0.5^2,\dots,0.5^2)$.
Since the model is linear and the prior is Gaussian, the posterior probability distribution is also Gaussian and we can solve for the posterior mean (Equation~(\ref{eq:linear_solve_mean})) and covariance (Equation~(\ref{eq:linear_solve_cov})) analytically.
In the hybrid pixel model we place a positive exponential prior on the pixels which are defined on a Mollweide grid. 
The purpose of the exponential prior is to favor sparser solutions for the map because the exponential distribution pushes most pixels towards zero intensity.
We use four times as many pixels as spherical harmonics.
Although the model is also linear in this case, the posterior distribution for the pixels is not analytic because of the non-Gaussian prior so we sample the posterior using MCMC instead.
As the end product of inference we save the spherical harmonic coefficients $\mathbf{y}=\mathbf{P}^\dagger\mathbf{p}$ rather than the pixels themselves.
The coefficients $\mathbf{y}$ can then be used to evaluate the map on a pixelated grid of arbitrary resolution via the matrix $\mathbf{P}$. 

To ensure that the difference in the inferred maps is not in part due to a difference in the scale of the priors, we take 5000 samples from the prior on $y$ and evaluate $\mathbf{P}\,\mathbf{y}$ for each; we then compute the standard deviation of these pixels and use that as the scale parameter in the exponential prior.
We implement the pixel model in the probabilistic programming language \textsf{numpyro} \citep{phan2019a} which is built on top of the \textsf{JAX} \citep{jax2018github} library, and we fit it using Hamiltonian Monte Carlo with the No-U-Turn-Sampler (NUTS) \citep{hoffman2014}.
\textsf{JAX} is a \textsf{numpy} like library which supports automatic differentiation, parallelization and GPUs.
We run the chains for 1000 tuning steps and 2000 final steps, monitoring divergences \citep{betancourt2013} and the R-hat  diagnostic  \citep{gelman1992a} to check for convergence.
Since all models we fit in this paper have hundreds if not thousands of parameters, it would be extremely challenging to sample the posterior  without the use of automatic differentiation and Hamiltonian Monte Carlo (at least for non-Gaussian priors).

\begin{figure}[t!]
    \begin{centering}
    \includegraphics[width=1.\linewidth]{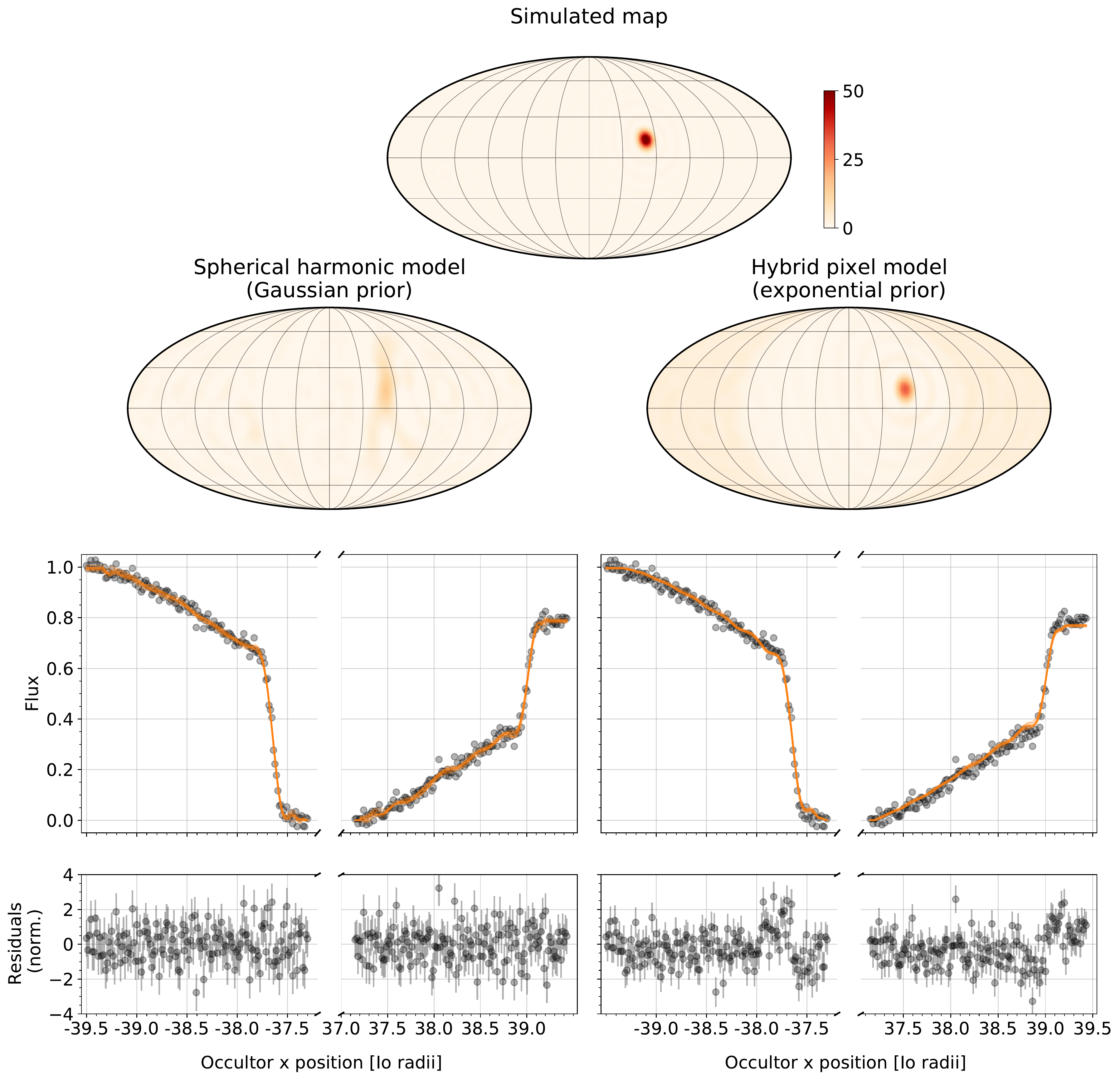}
    \oscaption{pixels_vs_harmonics}{%
        Comparison between two models fitted to a simulated light curve of an occultation by Jupiter which was generated from a map consisting of a single bright spot (top map).  
       Each of the columns beneath the simulated map shows the (median) posterior estimate of the map (top), the data and the posterior flux samples (middle row) and the residuals (bottom row).
        The left column corresponds to a model in which we place a Gaussian prior on spherical harmonic coefficients $\mathbf{y}$ and fit for $\mathbf{y}$ while the right column correspond to a hybrid model in which we fit for pixels $\mathbf{p}$ (with an exponential prior on pixel intensity) but we use $\mathbf{y}$ to compute the flux analytically with \textsf{starry}.
        The benefit of using pixels is that it is much easier to encode assumptions on what the map should look like in pixel space rather than spherical harmonic space.
        With the pixel model we are able to recover the simulated map precisely while the spherical harmonic model fails to do that. 
               \label{fig:pixels_vs_harmonics}
    }
    \end{centering}
\end{figure}

To visualize the inferred maps we plot the heatmap of the median intensity at each point on the map computed from posterior samples.
Results are shown in Figure~\ref{fig:pixels_vs_harmonics}.
The top row shows the simulated map, beneath it we show the inferred maps in Mollweide projection for the two models (second row), the data and posterior flux samples (orange lines) and the residuals with respect to the median flux (bottom row).
The difference between the two models is striking. 
The left map has an elongated feature which does not resemble the spot in the simulated map while the map on the right is nearly identical (except for a difference in intensity) to the simulated map. 
We should emphasize here that the fact that the pixel model results in a spot-like map is mostly a consequence of the exponential prior which favors sparse solutions in pixel space.
When we compared the spherical harmonic model to the pixel model using a prior with a lighter tail such as a Half Gaussian (Gaussian truncated at zero), we obtained a more elongated feature similar to shown on the left map in Figure~\ref{fig:pixels_vs_harmonics} but the map was still noticeably less complex because the positivity constraint substantially reduces the space of maps which fit the data well. 
Thus, the benefit of using the pixel model makes it easy to impose the positivity constraint on the map but also other constraints such as sparsity.

An important issue with the map on the right is that there is a series of concentric rings around the spot which result in a wave-like pattern in the predicted flux and the residuals.
This ringing pattern arises because representing spot-like features requires constructive interference between different spherical harmonic modes inside the spot and destructive interference elsewhere.
The pattern is more pronounced when we fit a low resolution map (approximately $180/20=9^\circ$ in this case) and the model tries to represent a feature below the resolution of the map.

Ringing is also the reason why the inferred spot for the pixel model appears to be noticeably dimmer than the simulated spot.
There is non-negligible leakage of total flux from the spot into the rings surrounding it, meaning that if we were to integrate the map intensity over a region encompassing the brightest part of the inferred spot, it would be an underestimate of the total emitted flux from the true spot within the same area.
Ringing is also undesirable because we want to avoid situations in which the model uses the rings to explain the data instead of just placing a spot directly. 
For example, we find that in some cases when we fit low degree maps, the model would place a bright spot on the unobserved side of Io in order to produce a ringing artefact on the observed side to explain an increase or decrease in brightness in the light curve. 
Fortunately, convolving the map with a spatial smoothing filter prior to evaluating the flux fixes this issue. 
We describe how to apply the smoothing filter in the following section.

\subsection{Smoothing out spurious features}
\label{ssec:spurious_features}
To suppress ringing artefacts which appear around inferred spots such as the one shown in Figure~\ref{fig:pixels_vs_harmonics}, we apply a spatial smoothing filter to the spherical harmonic coefficients.
Mathematically, the filtering operation is a convolution between the map and some kernel function $B(\theta,\phi)$.
Assuming both the map and the kernel function are expanded in terms of spherical harmonics, the convolution operation is simply a multiplication between the two sets of spherical harmonic coefficients.
We use a Gaussian-like kernel function given by
\begin{equation}
    B(\theta)=\frac{\exp \left(-\theta^{2} / 2 \sigma_s^{2}\right)}{2 \pi \sigma_s^{2}}
    \quad,
\end{equation}
where $\sigma_s$ is a parameter which sets the characteristic scale of the smoothing.
This function can be expanded in terms of spherical harmonics as
\begin{equation}
    B(\theta)=\sum_{l=0}^{\infty}\left(\frac{2 l+1}{4 \pi}\right) B_{l} \,\mathcal{P}_{l}(\cos \theta)
    \quad,
\end{equation}
where $B_l$ are the spherical harmonic coefficients and $\mathcal{P}_l$ are the associated Legendre polynomials.
They depend only on $l$ because all nonzero $m$ modes vanish due to azimuthal symmetry.
For $\sigma_s\ll 1$, $B_l$can be approximated as \citep{seon2007,white1995}
\begin{equation}
    B_l\simeq \exp\left[-\frac{1}{2}l(l+1)\sigma_s^2\right]
\end{equation}
The effect of this filter is to exponentially suppress features on scales smaller than $l\sim \sigma_s^{-1}$.

\begin{figure}[t!]
    \begin{centering}
    \includegraphics[width=1.\linewidth]{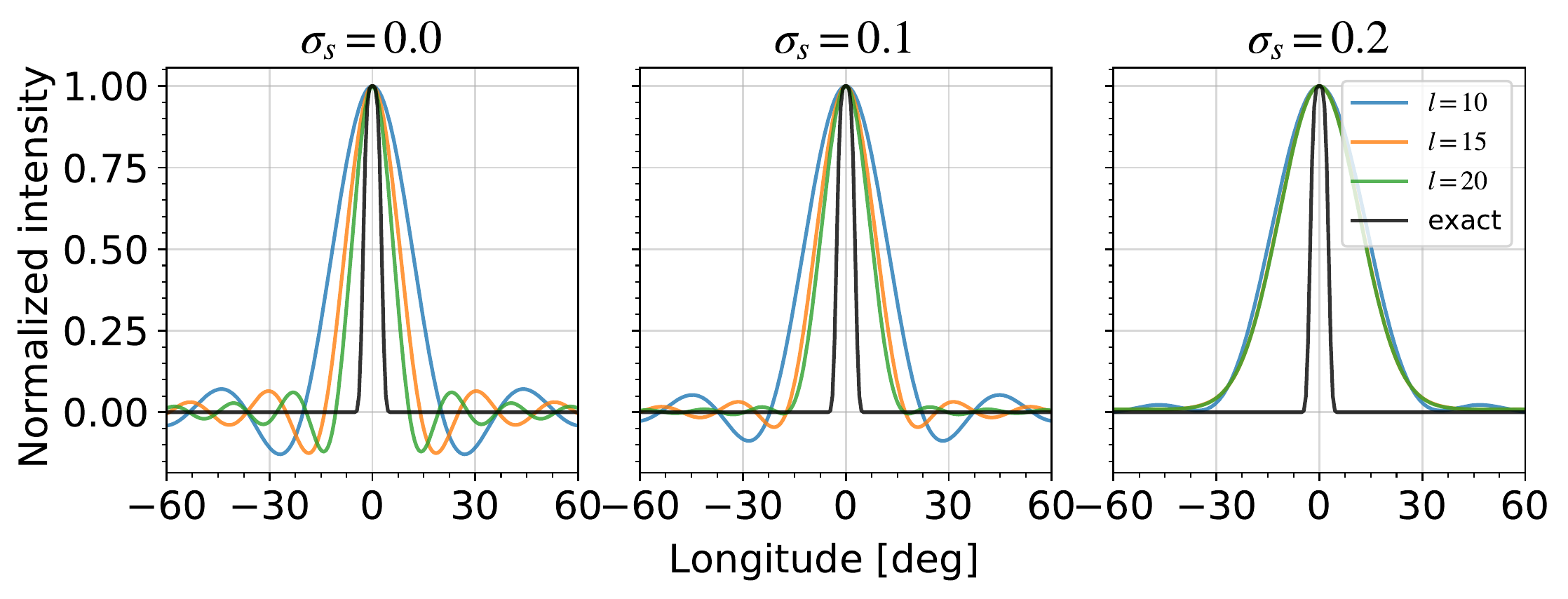}
    \oscaption{smoothing_kernel}{%
       Normalized intensity of a spherical harmonic expansion of 
        a Gaussian spot placed at $0^\circ$ latitude and longitude.
        The expansion is in the quantity $\cos\Delta\theta$ where $\Delta\theta=5^\circ$.
        The three plots show the spot profile for increasing values of the smoothing parameter 
    $\sigma_s=0$. 
        The colored lines correspond to spot expansions up to a certain order and the black line is the exact expansion.
        The purpose of smoothing is to taper higher order spherical harmonic coefficients in order to suppress ringing artefacts which result in negative intensities.
        High levels of smoothing suppress the ringing completely but as a result they increase the spot size and eliminate differences between expansions above a certain order.
        Intermediate levels of smoothing provide a compromise between the two extremes.
               \label{fig:smoothing_kernel}
    }
    \end{centering}
\end{figure}

Figure~\ref{fig:smoothing_kernel} shows the effect of (Gaussian) smoothing on a spherical harmonic expansion of a spot with a Gaussian intensity profile.
We place the spot at $0^\circ$ latitude and longitude and set the size of the spot $\Delta\theta$ to $5^\circ$.
All three panels show the exact profile of the spot (black line) and expansions up to three different orders $l$ (colored lines).
The panel on the left shows the expansion with no smoothing ($\sigma_s=0$) in which case the symmetric ringing around the center of the spot is clearly visible even at relatively high order ($l=20$).
The middle panel shows an intermediate level of smoothing with $\sigma_s=0.1$, meaning that all features on scales above $l\approx 10$ are exponentially suppressed.
The negative ringing is a lot less visible, albeit at the cost of having a slightly larger spot because suppressing higher order harmonics necessarily means that we lose some ability to represent smaller scale features.
For $\sigma_s=0.2$ (right) there is practically no ringing, but the expansions at $l=10$, $l=15$ and $l=20$ result in the spot of the same size because all coefficients above $l=5$ are significantly suppressed.

Thus, there is a trade-off between smoothing and the ability to resolve smaller scale features
in maps. 
In principle we can always get rid of ringing by fitting sufficiently high order maps.
In practice, the analytic integrals computed in \textsf{starry} become computationally unstable above $l\approx 20$ so instead of going to very high order we apply some smoothing to mitigate the ringing.
We find that setting $\sigma_s=2/l$ where $l$ is the order of expansion of the map  is a good default setting for $\sigma_s$.

\section{Results -- simulated data}
\label{sec:results_sim}
\subsection{Fitting simulated ingress/egress light curves}
\label{ssec:fitting_sim_ingress_egress}
In this section we generalize the example from Section~\ref{ssec:pixels_vs_harmonics} such that the simulated map includes an additional faint hot spot located at $-15^\circ$ N latitude and $-40^\circ$ E longitude with a diameter of $5^\circ$.
We set the amplitude of the ``faint'' spot to 30\% of the luminosity of the featureless map.
As before we assume that we know the geometry of the occultation exactly.
We generate the light curves from an $l=30$ map and fit an $l=20$ map and we apply a Gaussian smoothing filter to both the simulated map and the inferred map with the smoothing parameter set to $\sigma_s=2/l=0.1$.
We fit the model using NUTS with 1000 warm-up steps and 2000 final steps. 

We found that in cases where the simulated map consists of a bright and a faint spot the pixel model with an exponential prior such as the one we used in Section~\ref{ssec:pixels_vs_harmonics} does not recover the faint spot. 
Instead, we use a different heavy tailed prior called the Regularized Horseshoe prior\footnote{To be precise, we use the truncated version of the Regularized Horseshoe prior where the coefficients are required to be positive.} (also known as the ``Finnish Horseshoe'') \citep{piironen2017}.
This prior is specifically designed for use in Bayesian sparse linear regression. 
It is an improvement on the Horseshoe prior introduced in \cite{carvalho2010a}.
The key idea behind both kinds of Horseshoe priors is to set the scale for each regression coefficient (pixel) to a product of a global scale $\tau$ and a local scale $\lambda_i$ (where $i$ indexes all the pixels) and we marginalize over these scales by setting a prior for each. 
For clarity, we omit the discussion of  the Horseshoe priors here and refer the reader to Appendix~\ref{app:horseshoe}.

\begin{figure}[t!]
    \begin{centering}
    \includegraphics[width=1.\linewidth]{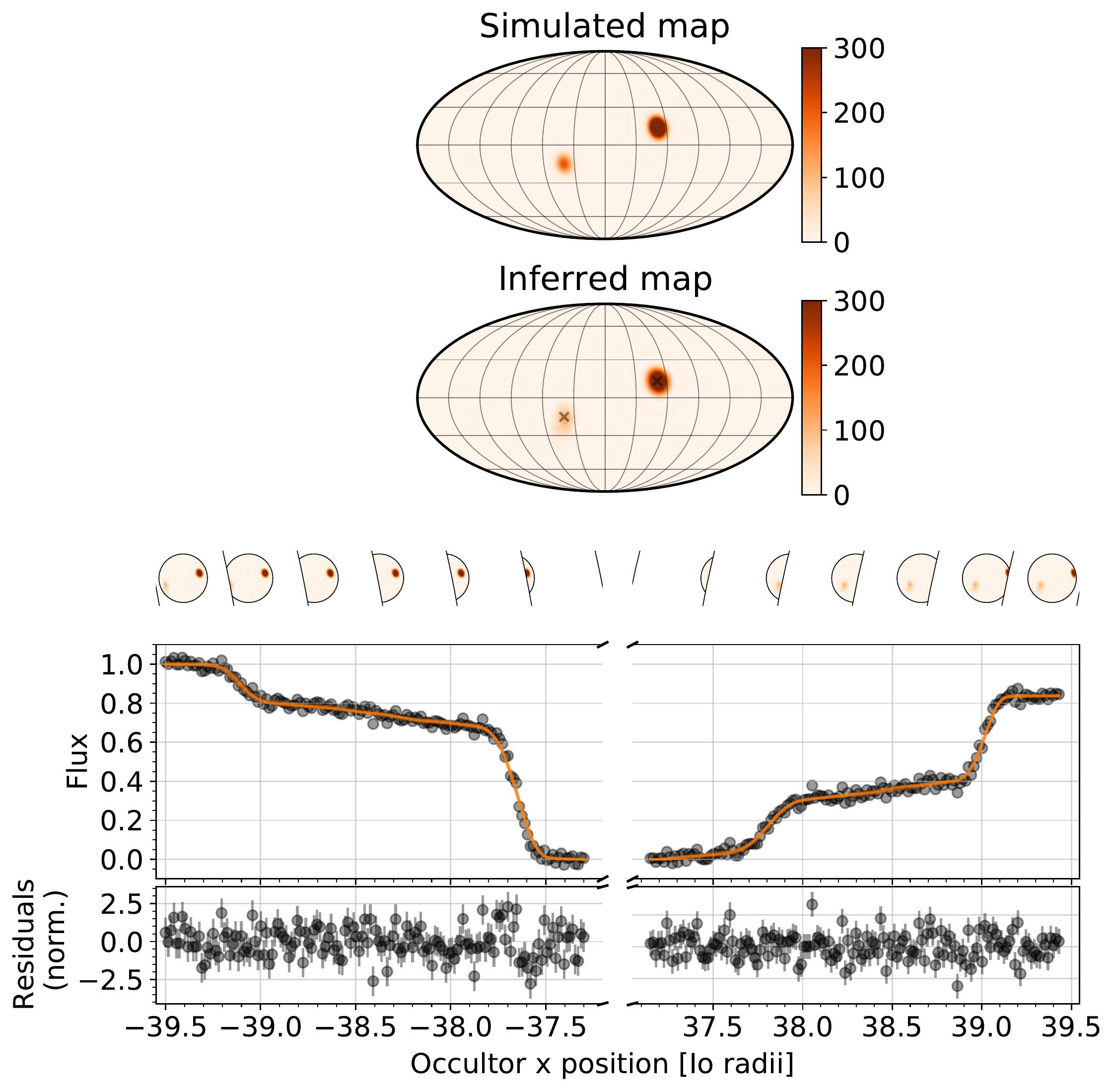}
    \oscaption{simulated_ingress_egress}{%
        Inferred $l=20$ map obtained by fitting a pair of simulated ingress/egress occultation light curves of Io using a hybrid pixel model with a Regularized Horseshoe prior on the pixels.
With this prior, the model is able to accurately recover the simulated map.
        The plot shows the simulated $l=20$ map (top), the posterior (median) estimate of the inferred map (second row), the inferred map as seen by the observer during the occultation (small circles), the data and posterior samples of the flux (orange lines), and the normalized residuals (bottom).
        The location of the simulated spots on the inferred map is marker with a grey cross (X).
       \label{fig:ingress_egress_sim_snr_50}
    }
    \end{centering}
\end{figure}

\begin{figure}[t!]
    \begin{centering}
        \includegraphics[width=1.\linewidth]{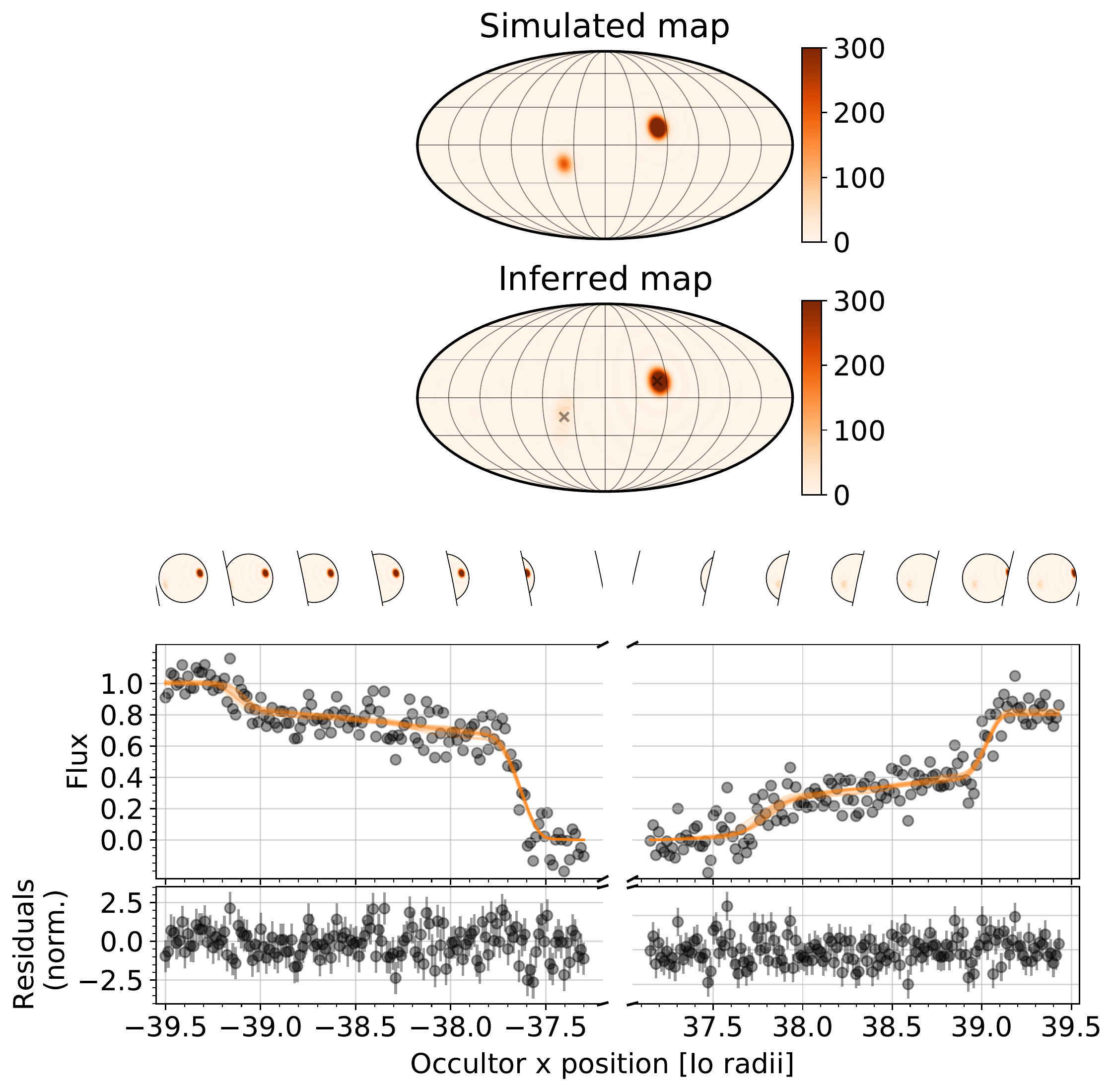}
    \oscaption{simulated_ingress_egress}{%
        Same as Fig.~\ref{fig:ingress_egress_sim_snr_50} except for light curves with SNR=10.
       \label{fig:ingress_egress_sim_snr_10}
    }
    \end{centering}
\end{figure}

The results for a high signal-to-noise light curve (SNR=50) are shown in Fig.~\ref{fig:ingress_egress_sim_snr_50}: the model recovers both spots.
The plot shows the simulated map (top row), the median posterior estimate of the inferred map (second row), the inferred map as seen by the observer during the occultation (small circles), the data and posterior samples of flux (orange lines) and the residuals with respect to a median estimate of the flux (bottom). 
The location of the simulated spots on the inferred map is marker with a grey cross (X).
The bright spot is nearly indistinguishable from the simulated spot; the fainter spot is somewhat less well constrained but the error in position for both is at most a few degrees.
Ringing artifacts are minimal because of the smoothing filter and there are no discernible patterns in the residuals.
We found that without the Horseshoe priors the model was not able to capture the first step in the light curve which is due to the fainter spot coming in our out of view; it would only recover the brighter spot.
In Fig.~\ref{fig:ingress_egress_sim_snr_10} we show the output of the same model assuming we have data of worse quality (SNR=10). 
In this case the model still recovers both spots but the error in position of the spots is greater.

\subsection{Comparison to a parametric model}
It is instructive to compare the model developed in the previous section to a parametric model in which we assume that the map contains some fixed number of spots and we fit for the positions, amplitudes and the sizes of the spots.
A parametric spot model might be useful if we want to fit for a small number of spots and we know what the map should look like.
To test how the parametrized spot model compares to the model defined in Section~\ref{ssec:fitting_sim_ingress_egress} we fit it to the light curves shown in Figure~\ref{fig:ingress_egress_sim_snr_50}. 
The model consists of 4 parameters: the latitude, longitude, amplitude and size of the spot.
We place uniform priors on the angles and positive Gaussian priors on other parameters. 
We find that if the number of modeled spots matches the number of simulated spots the model easily converges to the true solution. 
If the number of modeled spots is larger or smaller than the true number of spots the model does not converge to the true solution due to pathologies in the posterior distribution.

Although a parametric approach would likely have been sufficient for modeling the light curves shown in this work, it does not work well when the number of spots is unknown or if the features aren't spots.
There is no need to restrict ourselves to a parametric model because the pixel model gives the same results with only weak assumptions about the global structure of the map. 
The computational advantages of the parametric spot model relative to the pixel model are minimal. 
Even though the spot model has only 5 parameters per spot and the pixel model has thousands of parameters, the runtime for the pixel model is longer only by a factor of a few. 

\section{Results -- IRTF light curves}
\label{sec:results}
\subsection{1998 pair of occultations by Jupiter}
Having demonstrated that our model works well on simulated data, we turn to fitting observations of actual occultations of Io observed with the IRTF telescope.
We selected two pairs of high quality ingress/egress light curves relatively closely spaced in time so that the assumption that the surface map is identical for both the ingress and the egress light curve is at least approximately correct, although we allow for a difference in the overall amplitude of the maps between the two occultations.
We fit a pair of events from 1998, an ingress occultation observed on the 27th of August and an egress occultation observed  94 days later on the 29th of November.
As a reminder, the period of Loki's variability is around 540 days \citep{rathbun2002}.
For comparison, we also fit a pair of events observed nearly two decades later in 2017.

The predicted fluxes are given by 
\begin{align}
    \mathbf{f}_I&=\mathbf{A}_I'\,\mathbf{p}_I +b_I\,\mathbf{1}_{N_I}\\
    \mathbf{f}_E&=\mathbf{A}_E'\,\mathbf{p}_E +b_E\,\mathbf{1}_{N_E}
\end{align}
where $\mathbf{f}_I$ is the predicted flux for the ingress light curve with $N_I$ data points, $\mathbf{f}_E$ is the predicted flux for the egress light curve with $N_E$ data points, and $b_I$ and $b_E$ are constant flux offset parameters.
Since we assume that the map is the same for both light curves up to an overall amplitude difference, we have
\begin{equation}
    \mathbf{p}_E=a\,\mathbf{p}_I
    \quad,
\end{equation}
where $a$ is a dimensionless parameter.

The total log likelihood is the sum of the log likelihoods for individual light curves (Equation~(\ref{eq:likelihood})). 
To compute it we also need to specify the data covariance matrix defined in Equation~(\ref{eq:data_covariance_element}).
Each element of the covariance matrix consists of a white noise component (errobars) and a correlated noise component (Gaussian Process).
Since the IRTF light curves are not provided with estimated errobars, we choose to fit for all errobars simultaneously using a hierarchical approach in which we assign a global scale $l$ for all errorbars in a given light curve and we draw each individual errobar from a Half Normal distribution with standard deviation equal to that scale.
That is, we have 
\begin{align}
    \sigma_{i,I}&\sim \mathcal{N}^+(0, l_E),\quad i=1,\dots,N_I\\
    \sigma_{i,E}&\sim \mathcal{N}^+(0,l_I),\quad i=1,\dots, N_E
\end{align}
where $\sigma_{i,I}$ are the errobars for the ingress light curve and $\sigma_{i,E}$ are the errobars for the egress light curve.
The symbol $\mathcal{N}^+$ denotes a truncated Gaussian distribution restricted to positive values.
Treating errobars as free parameters means that we are introducing hundreds of more parameters (as many as there are data points) which is not an issue as long as these parameters are constrained by the data and our model is regularized.
The noise model as defined above is extremely flexible, it can account for a given feature in the light curve by inflating individual errobars (white noise) or by varying the characteristic timescale $\rho_\mathrm{GP}$ in the Mat\'ern 3/2 kernel.
All model parameters and their associated priors are listed in Table~\ref{tab:priors}. 

\renewcommand*{\arraystretch}{1.4}
\begin{table}[t!]
    \begin{center}
        \begin{longtable}{W{l}{4cm} W{l}{4cm} W{l}{6cm}}
            \label{tab:priors}
            \\
            \toprule
            \multicolumn{1}{c}{\textbf{Parameter(s)}}
             &
            \multicolumn{1}{c}{\textbf{Description}}
             &
            \multicolumn{1}{c}{\textbf{Prior}}
            \\
            \midrule
            \endhead
            \bottomrule                                 
            \\
            \caption{%
Parameters and priors for the final model. 
            The subscript $I$ denotes parameters associated with the ingress light curve while the subscript $E$ denotes parameters associated with the egress light curve. $t_{\mathrm{max},I}$ and  $t_{\mathrm{max},E}$ are the durations of the ingress and egress occultations respectively.
            The symbol $\mathcal{N}^+$ refers the normal distribution truncated at zero so that it only has support for positive values.
            }
            \endfoot
            $\tau$ & global pixel scale &  \begin{minipage}{0.5\textwidth}  \shortstack[l]{$\mathrm{Half}-\mathcal{C}(0,\tau_0)$, $\tau_0$ defined by\\Equation~(\ref{eq:horseshoe_tau0}) with $p_0=0.8D$}\end{minipage}
            \\
            $c^2$ & slab scale squared & $\mathrm{Inv}-\mathcal{G}(\frac{\nu}{2},\frac{
                \nu}{2}s^2)$, $s=1000$ and $\nu=4$
            \\
    $\overline{\lambda}_i , \quad m=1,\dots,D$  & local pixel scale & $\mathrm{Half}-\mathcal{C}(0,1)$
            \\
            $p_i, \quad i=1,\dots,D$& pixels & $\mathcal{N}^+(0, \tau\lambda_i)$, $\lambda_{i} =c \overline{\lambda}_{i}/\sqrt{c^{2}+\tau^{2} \overline{\lambda}_{i}^{2}}$
            \\
                $a$ & \begin{minipage}{0.2\textwidth}\shortstack[l]{relative change in \\map amplitude}\end{minipage}& $\mathcal{N}(1, 0.1)$
            \\
    $b_I$, $b_E$& flux offset &$\mathcal{N}(0, 4)$
            \\
            $l_I$, $l_E$& errorbar global scale&$\mathcal{N}^+(0,0.1)$
            \\
            $\sigma_{i,I}$, $\sigma_{i,E},\quad i=1,\dots,N$& individual errobars&
            $\mathcal{N}^+(0,l_I),\quad \mathcal{N}^+(0,l_E)$
            \\
            $\sigma_{\mathrm{GP}, I}$, $\sigma_{\mathrm{GP}, I}$ & GP standard deviation&$\mathcal{N}^+(0,0.1),\quad \mathcal{N}^+(0,0.1)$
            \\
            $\rho_{\mathrm{GP},I}$, $\rho_{\mathrm{GP},E}$ & GP timescale  & $\mathcal{N}^+(0, t_{max, I}),\quad  \mathcal{N}^+(0, t_{max, E})$
        \\
        \end{longtable}
    \end{center}
\end{table}

\begin{figure}[ht!]
    \begin{centering}
    \includegraphics[width=1.\linewidth]{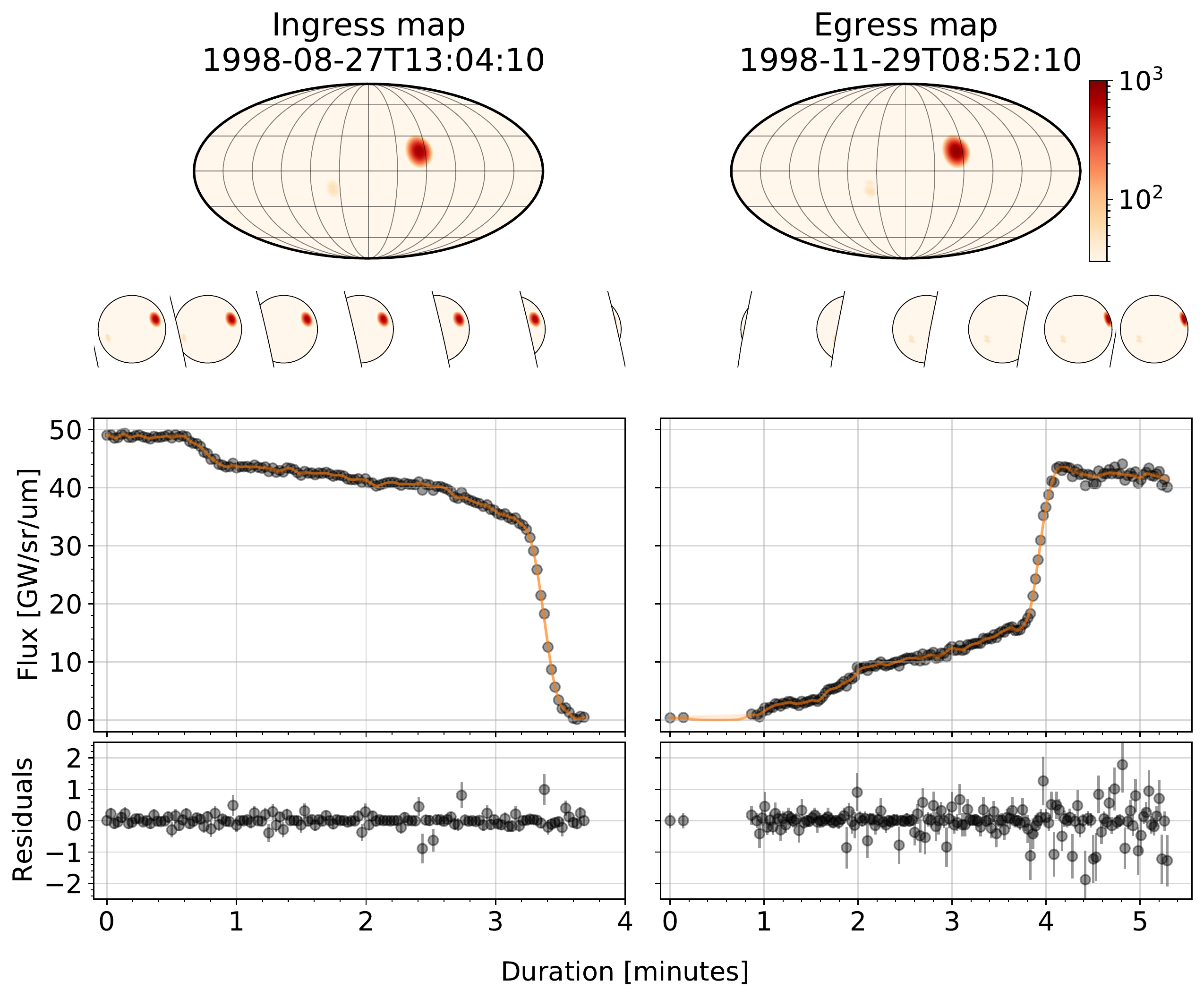}
    \oscaption{irtf_main_plots}{%
Inferred $l=20$ maps obtained by fitting a pair of observations of occultations of Io by Jupiter in 1998.
        The observations were made several months apart with the NASA Infrared Telescope Facility (IRTF).
We fit a single map to both observations simultaneously although we allow for a difference in the overall amplitude of the map between ingress and egress.
Our model includes a Gaussian Process which accounts for both the correlated noise due to seeing in the data, and the fact that our limited resolution map cannot fully capture the sharp steps in data.
We also fit for all errobars simultaneously using a hierarchical model and we plot the median estimates of those errobars.
        The plot shows  the inferred maps (top row), the same maps from the perspective of the observer during the occultation (small circles), the light curves and posterior samples of the flux including the Gaussian Process (orange lines), and the residuals with respect to a median flux estimate.
        The maps show two hotspots, the bright one is emission from Loki and the faint one is most likely emission from Kanehekili.
        A detailed view of the two hot spots is shown in Figure~\ref{fig:irtf_1998_spots}.
     \label{fig:irtf_1998}
    }
    \end{centering}
\end{figure}

\begin{figure}[t!]
    \begin{centering}
    \includegraphics[width=\linewidth]{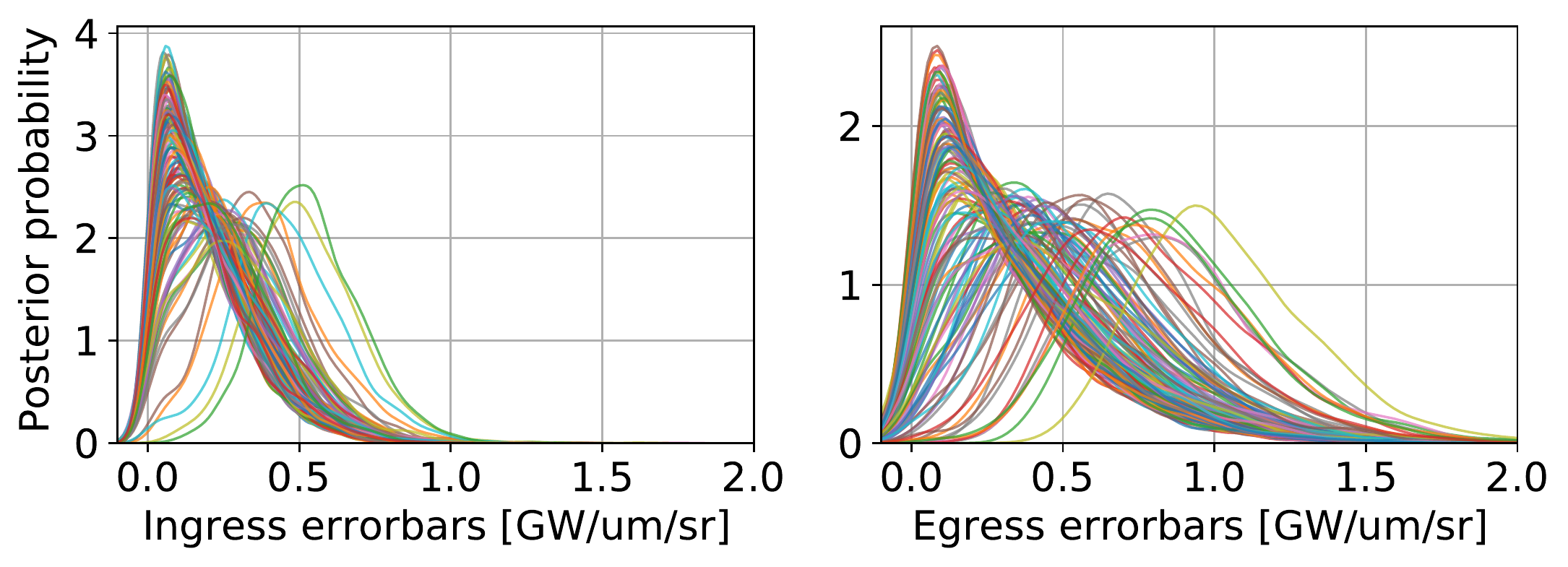}
    \oscaption{irtf_main_plots}{%
        Posterior distribution for the errobars of the 1998 pair of light curves shown in Figure~\ref{fig:irtf_1998}. 
        Each color corresponds to a single data point in the light curve.
        The errobars are estimated using a hierarchical model in which we assume  that each errobar is drawn from a Half Gaussian distribution with a global scale parameter common to all errobars in a single light curve.
       \label{fig:irtf_1998_errorbars}
    }
    \end{centering}
\end{figure}

To fit the model we sample the posterior using the NUTS sampler for 1000 warm-up steps and 3000 final steps and as before, we monitor divergences and the R-hat statistic to ensure convergence.
The inferred map for the 1998 pair of events is shown in Figure~\ref{fig:irtf_1998} and the median, 16th and 84th percentile estimates of model parameters from posterior samples are shown in Table~\ref{tab:irtf_1998}. 
The inferred map has two distinct hot spots, a bright hotspot in the Eastern hemisphere and a faint hotspot in the western hemisphere.
The orange lines in the second panels from the bottom are posterior samples of flux from the full model which includes the map model and the Gaussian Process.
The bottom panel shows the residuals with respect to the median flux.
Each data point is shown with the median prediction of its errorbar.
The posterior distribution for the errobars is shown in Figure~\ref{fig:irtf_1998_errorbars}.
Both the global scale for of the errobars and the individual errobars are well constrained by the data. 
The outlier points are naturally accounted for in our model because those points end up having higher variance.

Despite the fact that our noise model is extremely flexible, it does not seem to overfit the data because the two major steps in the light curves correspond to spots on the map.
The Gaussian Process accounts for the  variation in the data due to atmospheric variability, the emission due to surface features which are too faint to be well constrained by the physical model and the limitation of the physical model to capture the true size of the spot. 
There is a trade-off between the explanatory power of the noise model and the physical model. 
In general, we expect that a given feature in the light curve will be accounted for by the physical model if the physical model provides a better explanation for the data than the noise model.
We will return to this point shortly. 

\renewcommand*{\arraystretch}{1.4}
\begin{table}[t!]
    \begin{center}
        \begin{longtable}{W{l}{2cm} W{l}{5cm} W{l}{3cm} W{l}{3cm}}
            \label{tab:irtf_1998}
            \\
            \toprule
            \multicolumn{1}{c}{\textbf{Parameter}}
             &
            \multicolumn{1}{c}{\textbf{Description}}
            &
            \multicolumn{1}{c}{\textbf{Value}}
             &
            \multicolumn{1}{c}{\textbf{Unit}}
            \\
            \midrule
            \endhead
            \bottomrule                                 
            \\
            \caption{%
                Inferred parameters for the pair of occultations observed in 1998 using the IRTF telescope.
                }
            \endfoot
            $\tau$ & global pixel scale &   $0.870_{-0.221}^{+0.246}$ &  intensity
            \\
             $c$ & slab scale & $4304.302_{-1162.250}^{+2123.962}$ & intensity 
            \\
                $a$ & \begin{minipage}{0.2\textwidth}\shortstack[l]{relative change in \\map amplitude}\end{minipage} &  $1.154_{-0.013}^{+0.014}$ & dimensionless
            \\
            $b_I$ & flux offset ingress & $0.014_{-0.013}^{+0.077}$ & GW/um/sr
            \\
            $b_E$&flux offset egress & $0.362_{-0.335}^{+0.438}$ & GW/um/sr
            \\
            $l_I$& errorbar scale ingress & $0.253_{-0.030}^{+0.036}$ & GW/um/sr
            \\
            $l_E$& errorbar scale egress &$0.461_{-0.042}^{+0.042}$ & GW/um/sr
            \\
            $\sigma_{\mathrm{GP}, I}$ & GP standard deviation ingress & $0.633_{-0.042}^{+0.047}$ & GW/um/sr 
            \\
            $\sigma_{\mathrm{GP}, E}$ & GP standard deviation egress &$0.453_{-0.052}^{+0.054}$ & GW/um/sr
            \\
            $\rho_{\mathrm{GP},I}$ & GP timescale ingress & $0.078_{-0.010}^{+0.011}$ & minutes
            \\
            $\rho_{\mathrm{GP},E}$ & GP timescale egress & $0.099_{-0.021}^{+0.027}$ & minutes
            \\
        \end{longtable}
    \end{center}
\end{table}

To obtain some measure of the location of the inferred spots and its uncertainty, for each posterior sample of the spherical harmonic coefficients $\mathbf{y}$ we find the local maximum of intensity in a region around each of the spots, we then compute percentile estimates of spot latitude and longitude.
In addition to the locations of the spots, we also compute the total power emitted within a 15 degree range in latitude and longitude around the (inferred) center of the spots.
Both of these quantities are listed in Table~\ref{tab:irtf_1998_derived}. 

\begin{figure}[ht!]
    \begin{centering}
    \includegraphics[width=\linewidth]{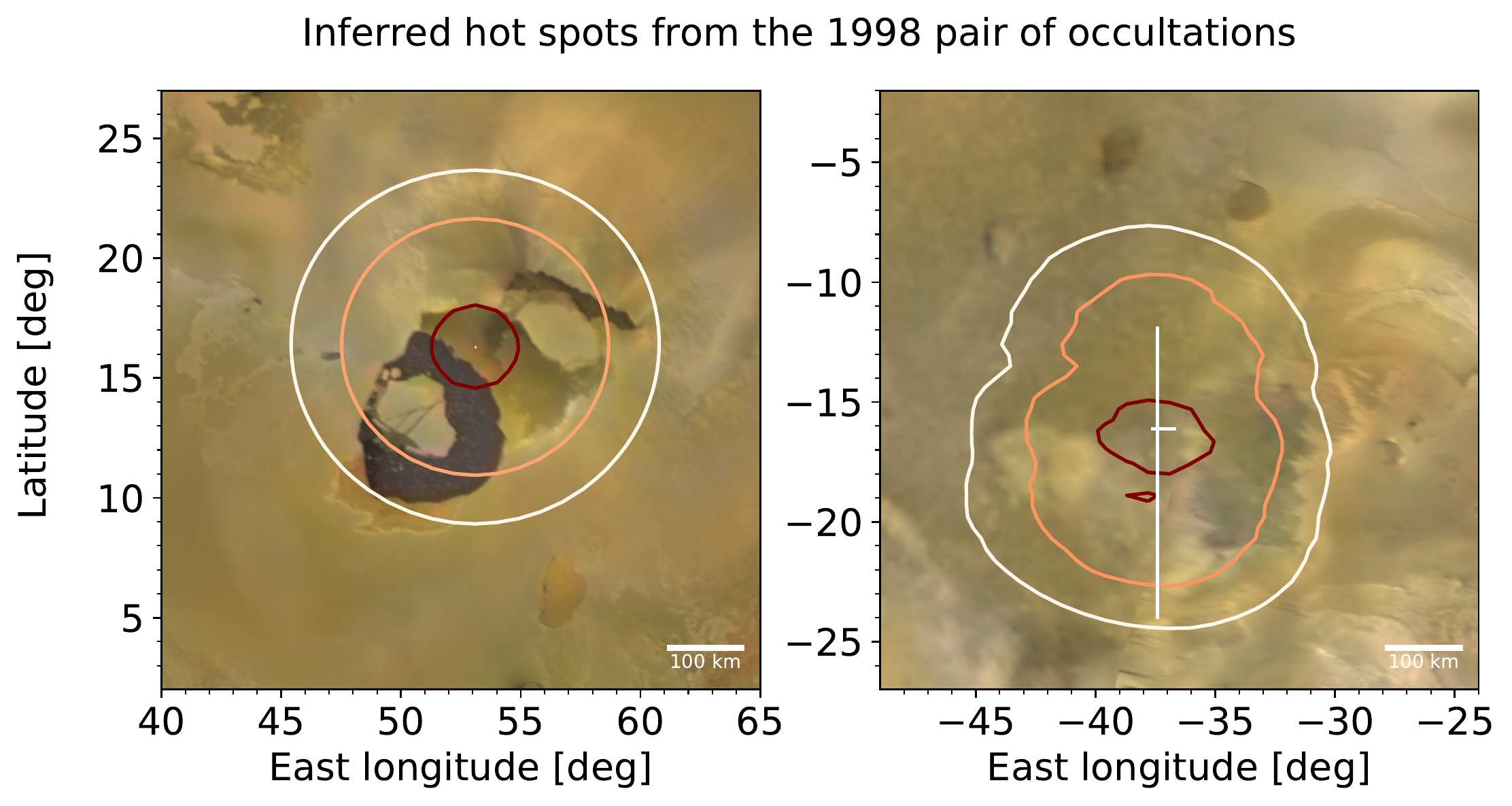}
    \oscaption{irtf_spots_plots}{%
        Contour plot of the hot spots shown in Figure~\ref{fig:irtf_1998} overlaid on top of the U.S. Geological Survey's map of the surface of Io which has been constructed from observations by the Galileo spacecraft.
        The left hotspot is centered around Loki Patera, the right hotspot is centered at the Kanehekili Fluctus lava flow.
The contour lines show the 5th, 50th, and 95th percentiles of intensity above an arbitrarily defined intensity of ``background'' region around the spot.
        \label{fig:irtf_1998_spots}
    }
    \end{centering}
\end{figure}

Figure~\ref{fig:irtf_1998_spots} shows  a contour map of both hot spots overlaid on top of the U.S. Geological Survey's map of the surface of Io \citep{williams2011} which was constructed from observations by the Galileo satellite.
The plot shows the error in the inferred latitude and longitude for each spot (white lines) and a contour map computed from the median posterior estimate of the map.
The contour lines correspond to the 5th, 50th, and 95th percentiles of intensity above an arbitrarily defined intensity of ``background'' region around the spot.

The location of the bright hotspot is $16.414_{-0.011}^{+0.027}\quad^\circ$ N latitude and $53.107_{-0.033}^{+0.041}\quad^\circ$ E longitude which corresponds to the location of Loki Patera.
The faint hot spot is at $-15.756_{-11.382}^{+3.812}\quad ^\circ$ N latitude and $321.568_{-4.208}^{2.803}\quad^\circ$ E longitude which is the Kanehekili Fluctus lava flow.
The error on the inferred position of peak intensity of the Loki hotspot is only a fraction of a degree which is much smaller than the minimum resolution of the map features set by the degree of the map; this corresponds to an uncertainty of less than a kilometer on the surface of Io.
Thus, \textbf{despite the fact that at $l=20$ the resolution of maps is limited to about $9^\circ$, we can constrain the centroid of the spots much more precisely}.
We should note however that this errorbar does not include a possible systematic shift in position due to a wrong estimate of Jupiter's effective radius (see Section~\ref{ssec:effective_radius}) or an error in the ephemeris data.

\renewcommand*{\arraystretch}{1.4}
\begin{table}[t!]
    \begin{center}
        \begin{longtable}{W{l}{4cm} W{l}{3cm} W{l}{3cm}}
            \label{tab:irtf_1998_derived}
            \\
            \toprule
            \multicolumn{1}{c}{\textbf{Parameter}}
             &
            \multicolumn{1}{c}{\textbf{Value}}
             &
            \multicolumn{1}{c}{\textbf{Unit}}
            \\
            \midrule
            \endhead
            \bottomrule                                 
            \\
            \caption{%
                Derived parameters of the two hot spots visible in Figure~\ref{fig:irtf_1998}.
                The latitude and longitude of the spots are estimated by finding the peak intensity of each spot for each posterior sample.
            The power of the spot is defined as the total emission from within a 15 degree circle around the (inferred) location of the spot.
        }
            \endfoot
            Spot 1 latitude  & $16.295^{+0.009}_{-0.010}$& degrees 
            \\
             Spot 1 (East) longitude & $53.105^{+0.011}_{-0.011}$& degrees 
            \\
                Spot 1 power  ingress & $52.394^{+0.898}_{-0.773}$ & GW/um 
            \\
                Spot 1 power  egress & $60.422_{-1.151}^{+1.194}$ & GW/um 
                \\
            Spot 2 latitude  & $-16.129_{-7.929}^{+4.268}$& degrees 
            \\
             Spot 2 (East) longitude & $-37.422_{-0.264}^{+0.778}$ & degrees 
            \\
                Spot 2 power  ingress & $5.156_{-0.634}^{+0.818}$ & GW/um 
                \\
                Spot 2 power  egress & $5.949_{-0.721}^{+0.957}$ & GW/um 
            \\
        \end{longtable}
    \end{center}
\end{table}

\begin{figure}[t!]
    \begin{centering}
    \includegraphics[width=1.\linewidth]{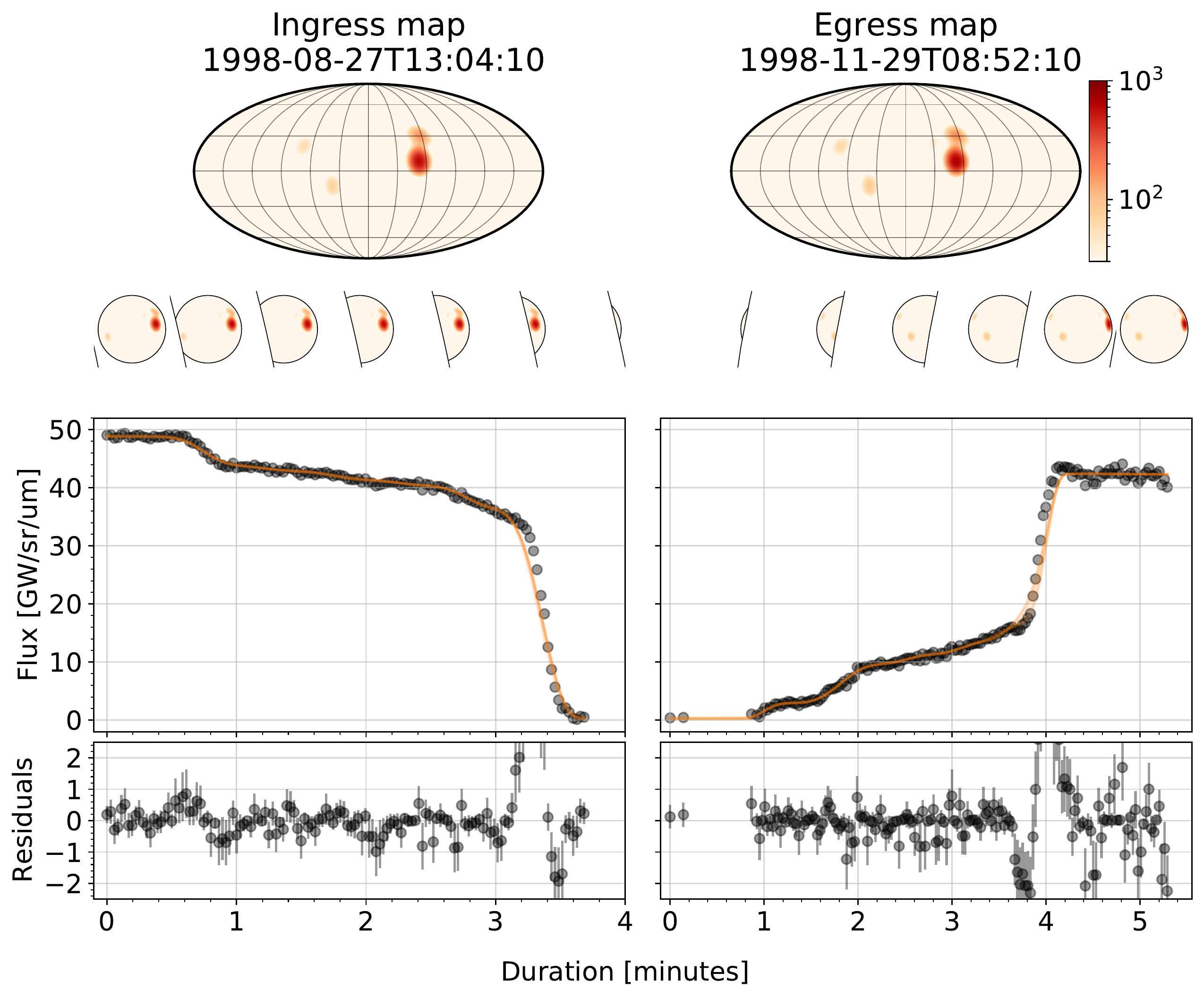}
    \oscaption{irtf_main_plots}{%
        Same as Figure~\ref{fig:irtf_1998} except this figure shows the output of a model which does not include a Gaussian Process to account for correlated structure in the data.
        As a result, the model tries to account for correlations in the data by placing two extra spots on the map and inflating errorbars.
        At least one of the two extra spots is artificial. 
     \label{fig:irtf_1998_no_GP}
    }
    \end{centering}
\end{figure}

To see how the above results change if we don't include a Gaussian Process in the noise model we fit the same pair of occultations from 1998 using the same priors, the results are shown in Figure~\ref{fig:irtf_1998_no_GP}.
The main difference in the inferred map compared to Figure~\ref{fig:irtf_1998} is that there are two extra spots visible in the map.
The two spots in the eastern hemisphere are the result of the model struggling to explain the main feature in the light curve at around 3.2 minutes in ingress light curve and 3.8 minutes in the egress light curve, which is due to  Loki coming out and into view during the occultation.
The model inflates the errobars around those times because it unable to make the spot small enough.
This results in the single hot spot at the location of Loki shown in Figure~\ref{fig:irtf_1998} morphing into two spots.
While this feature is almost certainly spurious, the other spot in the northwestern hemisphere appears more likely to be real.
Looking at the miniature maps in Figure~\ref{fig:irtf_1998_no_GP}, this spot is in view only at egress and it corresponds to a small step in the light curve starting at 0.9 minutes.
The estimated location of this spot is approximatel $\sim 20\quad^\circ$ N latitude and $\sim 69\quad^\circ$ W longitude.
This location corresponds to southern end of the mountain Mongibello Mons but there are no known persistent hotspots at that location so we cannot say if the hotspots is real or not without independently detecting it in other light curves.
Overall, we conclude that including the Gaussian Process prevents the physical model from overfitting the data although it might occasionally pick up a feature which is due to a faint real hot spot.
With a particular scientific goal in mind, it is straightforward to experiment with different noise models to test how the properties of any given spot change with different assumptions.

One other notable feature in Figure~\ref{fig:irtf_1998_no_GP} is that the first step in the ingress light curve at around 0.6 minutes isn't fully accounted for by the model.
Since the model had no trouble accounting for a similar a step for simulated data shown in Figure~\ref{fig:ingress_egress_sim_snr_50} the reason why it didn't do so in this case is likely because doing so would result in a poorer fit for the egress light curve.
The two occultations were observed months apart so we expect that our assumption that the map has not changed except for an overall amplitude is wrong in detail.

\subsection{2017 pair of occultations by Jupiter}
\begin{figure}[t!]
    \begin{centering}
    \includegraphics[width=1.\linewidth]{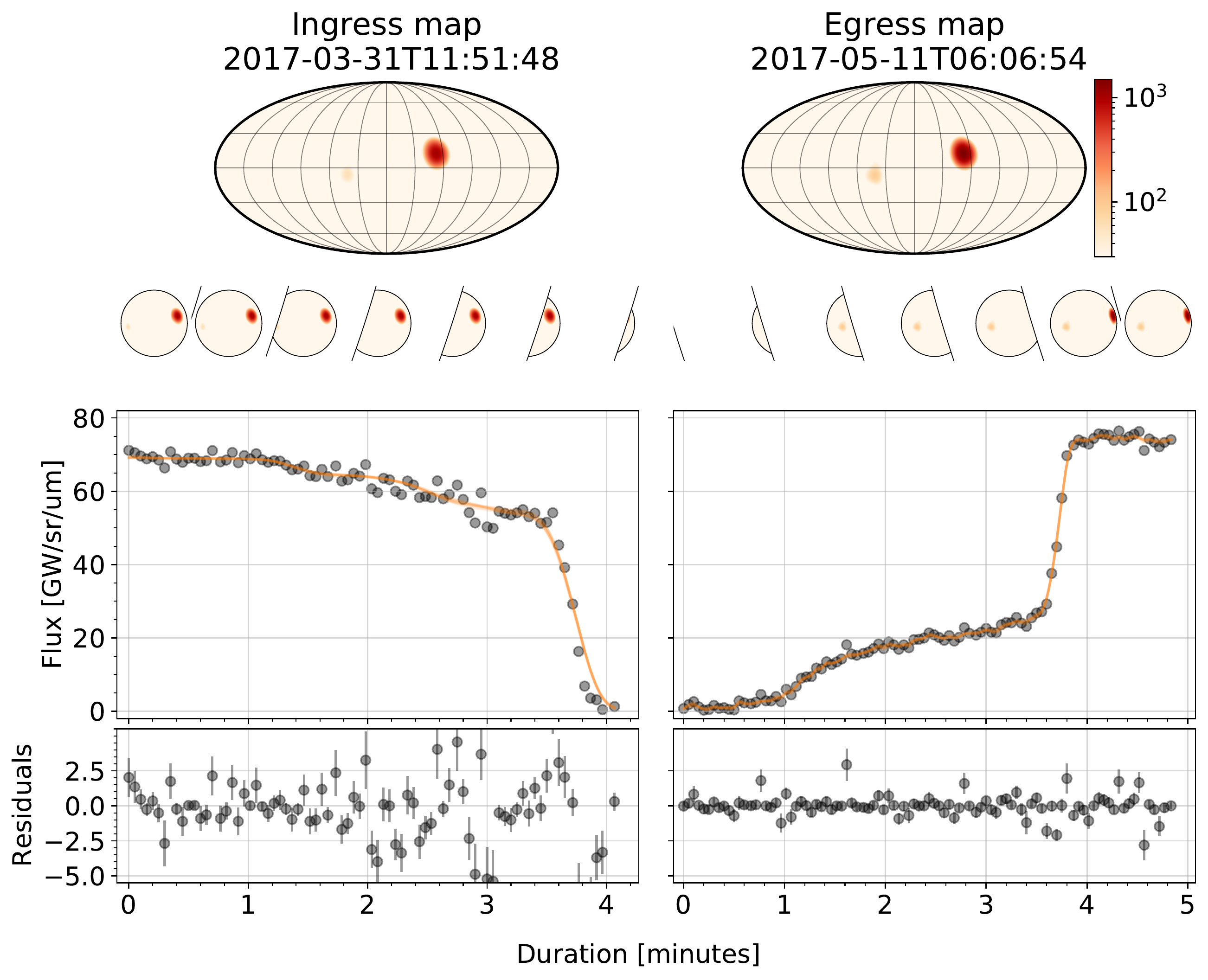}
    \oscaption{irtf_main_plots}{%
Inferred $l=20$ maps obtained by fitting a pair of observations of occultations of Io by Jupiter in 2017.
        The observations were made several months apart with the NASA Infrared Telescope Facility (IRTF).
We fit a single map to both observations simultaneously although we allow for a difference in the overall amplitude of the map between ingress and egress.
Our model includes a Gaussian Process which accounts for both the correlated noise due to seeing in the data, and the fact that our limited resolution map cannot fully capture the sharp steps in data.
We also fit for all errobars simultaneously using a hierarchical model and we plot the median estimates of those errobars.
        The plot shows  the inferred maps (top row), the same maps from the perspective of the observer during the occultation (small circles), the light curves and posterior samples of the flux including the Gaussian Process (orange lines), and the residuals with respect to a median flux estimate.
        The maps show two hotspots, the bright one is emission from Loki and the faint one is emission from Janus.
        A detailed view of the two hot spots is shown in Figure~\ref{fig:irtf_2017_spots}.
                \label{fig:irtf_2017}
    }
    \end{centering}
\end{figure}

\begin{figure}[t!]
    \begin{centering}
    \includegraphics[width=\linewidth]{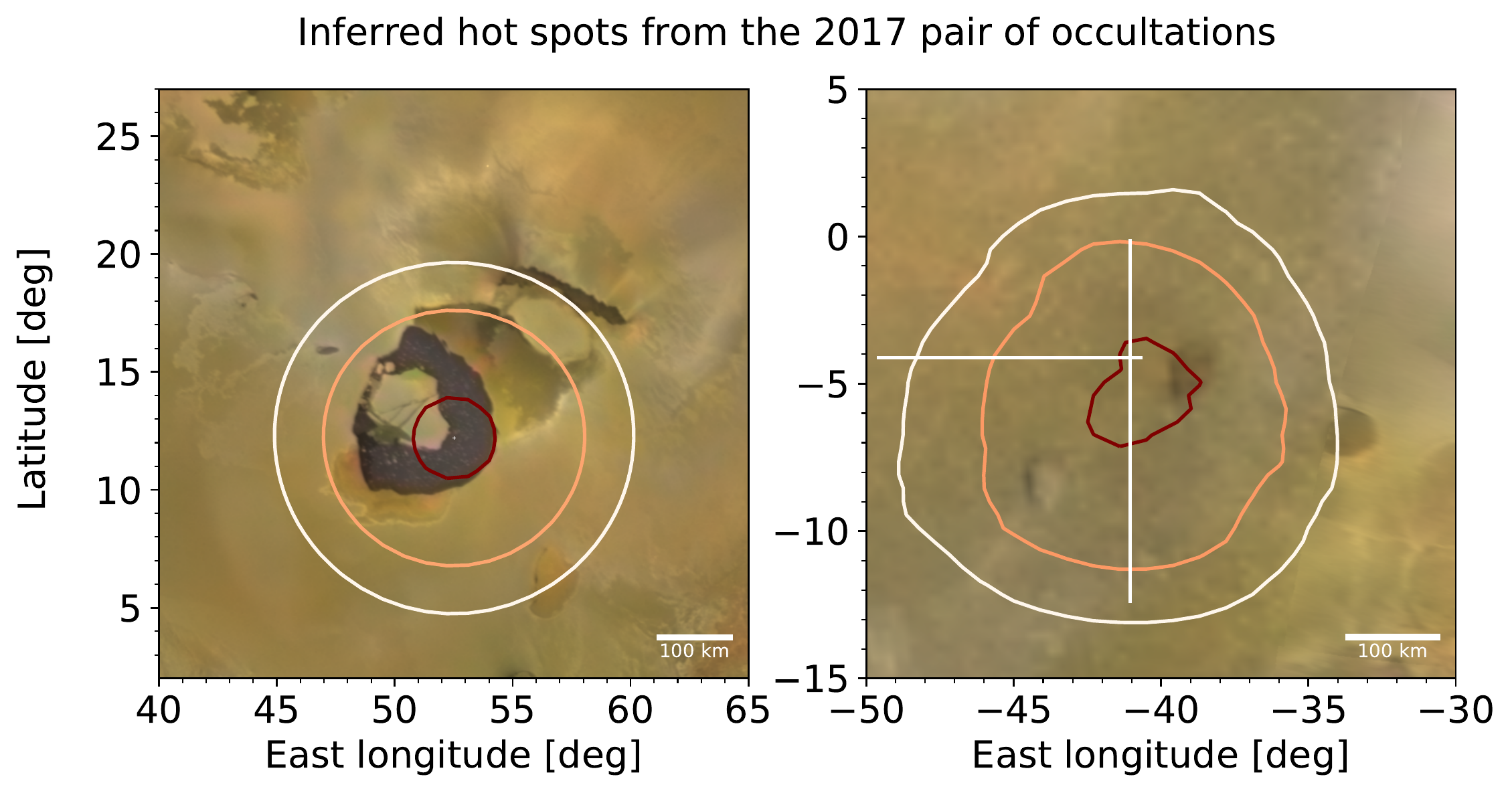}
    \oscaption{irtf_spots_plots}{%
        Contour plot of the hot spots shown in Figure~\ref{fig:irtf_2017} overlaid on top of the U.S. Geological Survey's map of the surface of Io which has been constructed from observations by the Galileo spacecraft.
        The left hotspot is centered around Loki Patera, the right hotspot is centered at Janus Patera.
The contour lines show the 5th, 50th, and 95th percentiles of intensity above an arbitrarily defined intensity of ``background'' region around the spot.
        \label{fig:irtf_2017_spots}
    }
    \end{centering}
\end{figure}

To test how our model generalizes to a different dataset we fit a pair of observations from 2017, an ingress occultation observed on the 31st of March and an egress occultation observed 41 days later on the 11th of May.
We use the exact same model as the one we used to produce Figure~\ref{fig:irtf_1998}.
The results are shown in Figure~\ref{fig:irtf_2017} and the inferred parameters are listed in Table~\ref{tab:irtf_2017}.
A contour plots of the two spots overlaid on top of a surface map of Io from Galileo observations is shown in Figure~\ref{fig:irtf_2017_spots} and the derived spot parameters are listed in Table~\ref{tab:irtf_2017_derived}.
The inferred map shows two spots, one of which is Loki. 
Figure~\ref{fig:irtf_2017_spots} shows that the peak emission from Loki is again constrained very precisely but the location of the peak appears to have shifted southward since 1998.
The location of the other hotspot corresponds to Janus Patera.

\renewcommand*{\arraystretch}{1.4}
\begin{table}[t!]
    \begin{center}
        \begin{longtable}{W{l}{2cm} W{l}{5cm} W{l}{3cm}  W{l}{3cm}}
            \label{tab:irtf_2017}
            \\
            \toprule
            \multicolumn{1}{c}{\textbf{Parameter}}
             &
            \multicolumn{1}{c}{\textbf{Description}}
            &
            \multicolumn{1}{c}{\textbf{Value}}
             &
            \multicolumn{1}{c}{\textbf{Unit}}
            \\
            \midrule
            \endhead
            \bottomrule                                 
            \\
            \caption{%
                Inferred parameters for the pair of occultations observed in 2017 using the IRTF telescope.
                }
            \endfoot
            $\tau$ & global pixel scale &   $0.973_{-0.247}^{+0.255}$ &  intensity
            \\
             $c$ & slab scale & $5866.860_{-1580.812}^{+2527.995}$ & intensity 
            \\
                $a$ &   \begin{minipage}{0.2\textwidth}\shortstack[l]{relative change in \\map amplitude}\end{minipage}  & $1.434_{-0.010}^{+0.011}$ & dimensionless
            \\
            $b_I$ & flux offset ingress & $0.025_{-0.024}^{+0.140}$ & GW/um/sr
            \\
            $b_E$& flux offset egress & $1.026_{-0.263}^{+0.238}$ & GW/um/sr
            \\
            $l_I$ & errobar scale ingress &$0.971_{-0.056}^{+0.056}$ & GW/um/sr
            \\
            $l_E$& errorbar scale egress & $0.520_{-0.093}^{+0.079}$ & GW/um/sr
            \\
            $\sigma_{\mathrm{GP}, I}$ & GP standard deviation ingress  & $0.126_{-0.055}^{+0.129}$ & GW/um/sr 
            \\
            $\sigma_{\mathrm{GP}, E}$ & GP standard deviation egress & $0.601_{-0.093}^{+0.083}$ & GW/um/sr
            \\
            $\rho_{\mathrm{GP},I}$ &  GP timescale ingress & $2.160_{-2.030}^{+3.333}$ & minutes
            \\
            $\rho_{\mathrm{GP},E}$ & GP timescale egress & $0.020_{-0.013}^{+0.011}$ & minutes
            \\
        \end{longtable}
    \end{center}
\end{table}

\renewcommand*{\arraystretch}{1.4}
\begin{table}[t!]
    \begin{center}
        \begin{longtable}{W{l}{4cm} W{l}{3cm} W{l}{3cm}}
            \label{tab:irtf_2017_derived}
            \\
            \toprule
            \multicolumn{1}{c}{\textbf{Parameter}}
             &
            \multicolumn{1}{c}{\textbf{Value}}
             &
            \multicolumn{1}{c}{\textbf{Unit}}
            \\
            \midrule
            \endhead
            \bottomrule                                 
            \\
            \caption{%
                Derived parameters of the two hot spots visible in Figure~\ref{fig:irtf_2017}.
                The latitude and longitude of the spots are derived by computing the centroid of points around the spots which are in the 90th percentile of intensity.
            The power is defined as the total emission from within a 15 degree circle around the (inferred) location of the spot.
        }
            \endfoot
                            Spot 1 latitude  & $12.193^{+0.011}_{-0.008}$ & degrees 
            \\
             Spot 1 (East) longitude & $52.515_{-0.009}^{+0.008}$ & degrees 
            \\
                Spot 1 power  ingress & $74.059_{-0.689}^{+0.911}$ & GW/um 
            \\
                Spot 1 power  egress & $106.158_{-1.412}^{+1.440}$ & GW/um 
                \\
            Spot 2 latitude  & $-4.114_{-8.327}^{+4.029}$ & degrees 
            \\
             Spot 2 (East) longitude & $-41.055_{-8.588}^{+0.433}$ & degrees 
            \\
                Spot 2 power  ingress & $5.171_{-1.007}^{+1.543}$ & GW/um 
                \\
                Spot 2 power  egress & $7.431_{-1.403}^{+2.227}$ & GW/um 
            \\
        \end{longtable}
    \end{center}
\end{table}

\section{Discussion}
\label{sec:discussion}
\subsection{Occultation mapping of Io}
We have presented a novel method for mapping the volcanic emission on Io from occultation light curves.
The method relies on the \textsf{starry} algorithm which enables fast analytic computation of occultation light curves by expanding the surface emission map in spherical harmonics.
Our method is different from past work because we do not assume we know where the volcanic features are located on the surface of Io, how many there are or what they look like.
Instead, we place weaker assumptions on the global structure of the map by requiring that the inferred map has positive intensity everywhere and most importantly, that it is \emph{sparse} \footnote{\cite{aizawa2020} also find the sparsity assumption useful in the context of exoplanet mapping.}.
Besides the model for the surface map, our method also incorporates a sophisticated noise model which includes a Gaussian Process and a hierarchical model for the (unknown) errorbars.
As a result of the sparsity and positivity constraints and the flexibility of the noise model,
our model is parsimonious: it places features on the map only if the data provides strong evidence for the existence of those features.
This property is not always desirable but it works very well for Io, a moon whose surface is covered with small, highly localized, and bright volcanic hotspots.
Our model is substantially more flexible than parametric methods which assume some fixed number of spots on the surface of Io and parametrize the spot properties because we do not need to make such strong assumptions about the surface.
The computational cost of the parametric models is comparable to our approach.

To test our method we first fit a simulated dataset and then observations of real occultations of Io by Jupiter, observed using NASA's Infrared Telescope Facility.
We choose two pairs of light curves to demonstrate that the model can recover known hotspots.
Each pair consists of an ingress observation and an egress observation of an occultation. 
The two observations are sufficient to break the degeneracy in the position of the spots because Jupiter's limb sweeps across the projected disc of Io at different angles at ingress and egress.
From the 1998 observations we infer a map consisting of two spots whose locations correspond to well known hotspots on the surface of Io, the major volcano Loki and the lava flow Kanehekili.
We also find circumstantial evidence for a third hot spot.
Because our model is fully probabilistic, we can derive uncertainties on the location of the inferred hot spots and the total flux emitted from each spot.
In addition to the pair of observations from 1998, we test the model on another pair of observations from 2017.
We again find two hotspots, one of which is Loki (although the peak of the intensity shifted relative to the 1998 map) and the other is Janus Patera.

The main limitation of our model besides the fact that we do not yet account for time-dependent maps is the limited resolution of the spherical harmonic maps for which \textsf{starry} can compute occultation light curves.
Although we can still constrain the peak intensity of hotspots (or some other measure of the center of emission) with much higher precision, we lose the ability to resolve two spots close to each other and the ability to constrain the actual size of the spots. 
It is quite possible that there is a way around this problem if we compute the various integrals in \textsf{starry} numerically using high precision arithmetic.
Since all these operations are contained in the design matrix $\mathbf{A}$ (assuming that the ephemeris is fixed) we would only need to do the expensive computation once for each occultation.
To be able to constrain the size of the Loki hotspot for example, we would need to fit maps on the order of $l\approx 50$.
However, the kind of detailed mapping of Loki's magma lake with a precision of a few kilometers done by \cite{dekleer2017} will not be possible with a spherical harmonic model because it would require maps of extremely large degrees ($l\gtrapprox 700$).

\subsection{Relevance to mapping of exoplanets} 
\label{ssec:discussion_exoplanets}
Hot Super Earths with (probably) molten surfaces such as 55 Cancri e are prime targets for future observations with the James Webb Space Telescope (JWST) \citep{samuel2014,henning2018}.
The model we have presented in this paper can easily be applied to observations of exoplanets. 
In fact, one of the motivations for this work was to test methods developed largely for the purpose of mapping exoplanets in a context where the ground truth is more easily accessible because Io is in the Solar System.
To apply our model to exoplanet observations we just need to compute the design matrix $\mathbf{A}$ given a specification of the planet's orbit which is very straightforward in \textsf{starry}
and since the observations of the secondary eclipses (occultations) of exoplanets will never be as high quality as observations of Io, the limited resolution of spherical harmonic maps is more than sufficient for modeling exoplanet observations.
The question of what kind of features on the surfaces of volcanic exoplanets it will be possible to constrain with secondary eclipse observations using JWST and with what precision is one we aim to address in future work.

Besides being useful for modeling secondary eclipses of volcanic Super-Earths, our model can be used for modeling surfaces of gaseous exoplanets with sparse features.
These will be much easier to observe with JWST  than Super-Earths because of their larger size and higher surface temperatures.
Even if the sparsity assumption is not valid for these types of planets the hybrid pixel/spherical harmonic model can be easily be used with a different set of priors more appropriate for these observations.
For example, a sensible prior for the surface of a fast rotating gaseous exoplanet might be requiring some degree of azimuthal symmetry for the inferred map.

\subsection{Future work}
This paper is a first step in a series of papers dedicated to probabilistic modeling of Io's surface. 
A major issue with the model presented in this work is that we cannot naturally account for the time variability of surface emission.
In Paper II in this series, we aim to fit an ensemble of light curves (occultations by Jupiter, mutual occultations and phase curves) with a generalized model which will enable us to infer a time-dependent map of the entire surface and quantify the time variability over time scales of decades.
A spatio-temporal map of volcanic emission spanning such a long timescale will help us better understand the global evolution of volcanism on Io. 
In addition to modeling global properties of volcanism of Io, it would also be interesting to constrain the time evolution of peak emission in Loki Patera with high accuracy in order to better understand the resurfacing process.

Finally, it should be straightforward to extend our model for use in fitting \emph{resolved} observations of Io (either occulted or not) such as the adaptive optics observations done by \cite{dekleer2016a} and \cite{dekleer2016}.
Since we would not need to compute integrated flux over complicated boundaries in that case, we could fit maps of much higher order.

\vspace{+3em}

We would like to thank Will Farr, Katherine de Kleer, David W. Hogg, and the Astronomical Data Group at the Center for Computational Astrophysics for their help and thought-provoking discussions. Fran Bartoli\'c acknowledges the support and funding from the 2020 Flatiron Institute Center for Computational Astrophysics Pre-Doctoral program which made this paper possible.

% Bibliography 
\bibliography{bib}

\appendix
\section{Horseshoe priors}
\label{app:horseshoe}
The Regularized Horseshoe prior \citep{piironen2017} is specifically designed for use in Bayesian sparse linear regression. 
It is a generalization of the Horseshoe prior introduced in \cite{carvalho2010a}.
The idea behind the Horseshoe prior is to set the scale for each regression coefficient (pixel) to a product of a global scale $\tau$ and a local scale $\lambda_i$ where $i$ indexes all the pixels.
The Horseshoe prior is defined hierarchically as 
\begin{equation}
\begin{aligned}
    &\tau  \sim \mathrm{Half}-\mathcal{C} \left(0, \tau_{0}\right)\\
    &\lambda_{i}  \sim \mathrm{Half}-\mathcal{C} (0,1) \\
    &p_{i}  \sim \mathcal{N}\left(0, \tau \lambda_{i}\right) 
\end{aligned}
    \quad,
    \label{eq:horseshoe}
\end{equation}
where $p_i$ are the pixels.
Each local scale parameter is drawn from a unit scale heavy tailed Half Cauchy distribution which allows for very large values of the pixels.
The global scale parameter is also a free parameter, drawn from a Half Cauchy distribution with the scale equal to $\tau_0$.
The Horseshoe prior is closely related to the spike-and-slab prior which is a mixture between a delta function prior at zero (\emph{spike}) and some other prior elsewhere (\emph{slab}).

The Regularized Horseshoe prior adds another level to Equation~(\ref{eq:horseshoe}) in order to allow fine tuned control of sparsity and to regularize very large values of coefficients in cases where the data is only weakly constraining.
\cite{piironen2017} show that the Regularized Horseshoe prior can be considered as a continuous counterpart of the spike-and-slab prior with a finite slab width $c$ whereas the Horseshoe prior resembles a spike-and-slab prior with a slab of infinite width.
The prior is defined by
\begin{equation}
\begin{aligned}
    &\tau  \sim \mathrm{Half}-\mathcal{C}\left(0, \tau_{0}\right)\\
    &c^{2}  \sim \mathrm{Inv}-\mathcal{G}\left(\frac{\nu}{2}, \frac{\nu}{2} s^{2}\right) \\
    &\overline{\lambda}_{i}  \sim \mathrm{Half}-\mathcal{C}(0,1)\\
    &\lambda_{i} =\frac{c \overline{\lambda}_{i}}{\sqrt{c^{2}+\tau^{2} \overline{\lambda}_{i}^{2}}} \\
    &p_{i}  \sim \mathcal{N}^+\left(0, \tau \lambda_{i}\right) 
\end{aligned}
    \label{eq:reg_horseshoe}
\end{equation}
Integrating out the slab scale $c$ implies a marginal $\mathrm{Student}-\mathit{t}\,(\nu,0,s)$ prior for pixels far from zero.
When pixels $p_i$ are close to zero ($\tau^2\overline{\lambda}_i^2\ll c^2$) we have $\lambda_i^2\rightarrow\overline{\lambda}_i^2$ and the prior approaches the original Horseshoe.
When pixels are far from zero ($\tau^2\overline{\lambda}_i^2\gg c^2$) then $\lambda_i^2\rightarrow c^2/\tau^2$ and the prior approaches $\mathcal{N}(0, c)$.

\cite{piironen2017} suggest the following expression to set the scale parameter $\tau_0$ which is an estimate of the global scale of the pixels 
\begin{equation}
    \tau_0=\frac{p_0}{D-p_0}\frac{\sigma}{\sqrt{n}}
    \quad,
    \label{eq:horseshoe_tau0}
\end{equation}
where $p_0$ is our prior guess for the number of significant pixels that are sufficiently far above zero, $D$ is the total number of pixels, $n$ is the number of data points and $\sigma$ is the standard deviation of the data points (the errobars).
Thus, we only need to specify $p_0$, the degree of freedom parameter $\nu$ and the slab width $c$.

When using the Regularized Horseshoe prior in a small data regime it is often necessary to use the non-centered parametrization to avoid funnels in the posterior which are often present in hierarchical models\footnote{\url{https://mc-stan.org/docs/2_26/stan-users-guide/reparameterization-section.html}}.
The purpose of this reparametrization is to reduce the dependence between the hyperparameters in the posterior.
To implement the non-centered parametrization we replace priors in Equation~\ref{eq:reg_horseshoe} with zero mean and unit variance priors and rescale them with deterministic transforms as follows
\begin{equation}
\begin{aligned}
    &\overline{\tau} \sim \mathrm{Half}-\mathcal{C}\left(0, 1\right)\\
    &\overline{c^{2}}  \sim \mathrm{Inv}-\mathcal{G}\left(\frac{\nu}{2}, 1\right) \\
    &\overline{\lambda}_{i}  \sim \mathrm{Half}-\mathcal{C}(0,1) \\
    &\overline{p}_{i}  \sim \mathcal{N}^+\left(0, 1\right)\\
    &\tau=\tau_0\overline{\tau}\\
    &c^2=\frac{\nu}{2}s^2\,\overline{c^2}\\
    &\lambda_{i} =\frac{c \overline{\lambda}_{i}}{\sqrt{c^{2}+\tau^{2} \overline{\lambda}_{i}^{2}}} \\
    &p_i = \tau\lambda_i\,\overline{p}_i
\end{aligned}
    \label{eq:reg_horseshoe_noncentered}
\end{equation}
We find that without using the non-centered parametrization the sampling is problematic and there are many divergences in the gradients of the parameters; with the non-centered parametrization there are no problems with sampling.

\clearpage

\end{document}